\documentclass{mn2e}
\input psfig.sty

\newcommand{\g}{$\gamma$}
\newcommand{\msun}{{{\rm M}_{\sun}}}
\newcommand{\xte}{{\it RXTE}}
\newcommand{\ginga}{{\it Ginga}}
\newcommand{\asca}{{\it ASCA}}
\newcommand{\gro}{{\it CGRO}}
\newcommand{\sax}{{\it Beppo\-SAX}}
\newcommand{\ledd}{L_{\rm E}}
\newcommand{\source}{GX~339--4}

\title{\source: the distance, state transitions, hysteresis and spectral correlations}

\author[A. Zdziarski et al.]
{Andrzej A.~Zdziarski,$^1$ Marek Gierli\'nski,$^{2,3}$ Joanna Miko{\l}ajewska,$^1$ \newauthor 
Grzegorz Wardzi\'nski,$^1$ David M. Smith,$^4$ B.~Alan~Harmon$^5$ and Shunji Kitamoto$^6$ 
\\
$^1$Centrum Astronomiczne im.\ M. Kopernika, Bartycka 18, 00-716 Warszawa, Poland; aaz@camk.edu.pl \\
$^2$Department of Physics, University of Durham, Durham DH1~3LE, UK\\
$^3$Obserwatorium Astronomiczne Uniwersytetu Jagiello\'nskiego, Orla 171, 30-244 Krak{\'o}w, Poland\\
$^4$Physics Department, University of California Santa Cruz, Santa Cruz, CA 95064, USA\\
$^5$NASA Marshall Space Flight Center, SD50, Huntsville, AL 35812, USA\\
$^6$Physics Department, College of Science, Rikkyo University
3-34-1, Nishi-Ikebukuro, Toshima-ku, Tokyo, 171-8501, Japan\\
}
\date{Accepted 2004 March 12. Received 2003 December 15}

\pagerange{\pageref{firstpage}--\pageref{lastpage}}
\pubyear{2004}

\begin{document}

\maketitle

\label{firstpage}

\begin{abstract} We study X-ray and variability and distance of \source. We 
derive the distance of $\ga 7$ kpc, based on recent determination of the 
binary parameters. We study data from the ASM aboard \ginga, the BATSE aboard 
\gro, and the ASM, PCA and HEXTE aboard \xte. From 1987 to 2004, \source\ 
underwent $\sim$15 outbursts and went through all known states of black-hole 
binaries. For the first time, we present the PCA data from the initial hard 
state of the outburst of 2004. We then study colour-colour and colour-flux 
correlations. In the hard state, there is a strong anticorrelation between the 
1.5--5 and 3--12 keV spectral slopes, which we explain by thermal Comptonization 
of disc photons. There is also a softening of the spectrum above 3 keV with the 
increasing flux that becomes stronger with increasing energy up to $\sim$200 
keV. This indicates an anticorrelation between the electron temperature and 
luminosity, explained by hot accretion models. In addition, we see a variable 
broad-band slope with a pivot at $\sim$200 keV. In the soft state, there is a 
high energy tail with varying amplitude beyond a strong and variable blackbody 
component. We confirm the presence of pronounced hysteresis, with the 
hard-to-soft state transitions occurring at much higher (and variable) 
luminosities than the soft-to-hard transitions. We fit the \xte/ASM data with a 
model consisting of an outer accretion disc and a hot inner flow. State 
transitions are associated then with variations in the disc truncation radius, 
which we fit as $\sim 6GM/c^2$ in the soft state and several times that in the 
hard state.  The disappearence of the inner disc takes place at a lower 
accretion rate than its initial appearance due to the dependence of the 
transitions on the source history. We provide further evidence against the X-ray 
emission in the hard state being nonthermal synchrotron, and explain the 
observed radio-X-ray correlation by the jet power being correlated with the 
accretion power. \end{abstract}


\section{Introduction}
\label{s:intro}

The Galactic X-ray binary \source\ was discovered $>$30 years ago (Markert et 
al.\ 1973), and it has been extensively studied since then. Its optical 
companion is undetectable and most of the observed optical emission originates 
in an accretion disc. The system is classified as a low-mass X-ray binary from 
upper limits on the luminosity of the companion star (e.g., Shahbaz, Fender \& 
Charles 2001). 

\source\ is a transient source exhibiting state transitions between soft, hard 
and off states. Based on its temporal and spectral characteristics, it is 
considered to be a black hole binary (Sunyaev \& Revnivtsev 2000; Zdziarski et 
al.\ 1998, hereafter Z98), which is in agreement with an estimate of its mass 
function $\ga 2\msun$ at the 95 per cent confidence (Hynes et al.\ 2003). 

Most of its X$\gamma$-ray spectra in luminous states are remarkably similar to 
those of Cyg X-1, and well fitted by Comptonization (Z98; Wardzi\'nski et al.\ 
2002, hereafter W02). However, while the two sources share common properties on 
time scales $\la$ a few days, their evolution patterns are rather different on 
longer time scales (e.g., Smith, Heindl \& Swank 2002b). In particular, the 
spectral evolution of Cyg X-1 is directly correlated with the luminosity 
changes, whereas the spectral changes in \source\ have been observed to lag the 
variations of the luminosity. These lags constitute a hysteresis in the 
lightcurve of the source; the hard-to-soft state transitions during the rise 
phase occur at higher luminosities than the soft-to-hard ones during the decline 
phase.

A major difference between \source\ and Cyg X-1 is that the latter is a 
persistent source in a high-mass X-ray binary. Thus, accretion in Cyg X-1 is via 
a focused wind from the high-mass companion, while in \source\ it is due to 
Roche-lobe overflow from the low-mass companion. The different behaviour of low 
and high-mass black-hole binaries is likely to be due to different disc sizes 
(e.g., Smith et al.\ 2002b). In a low-mass X-ray binary, the disc is expected to 
be large. This then leads to delays in the propagation of the accretion rate 
from the disc outer regions to the hot inner flow. Another consequence of the 
different types of accretion flows in the two systems is that whereas Cyg X-1 is 
a persistent source, in which the bolometric luminosity (and presumably the 
accretion rate) varies only by $\sim$5 or so (Zdziarski et al.\ 2002, hereafter 
Z02), \source\ is a transient, in which accretion is unstable and the accretion 
rate in the vicinity of the black hole varies by at least several orders of 
magnitude. Consequently, \source\ goes through a wider range of spectral states 
than Cyg X-1. 

In this paper, we first estimate the distance to the source based on recent 
optical/IR data. Then we study the long-term variability of \source\ during 
1987--2004 and its theoretical interpretation. We use data from the All-Sky 
Monitors (ASM) on board \ginga\/ and {\it Rossi X-ray Timing Explorer (RXTE)\/} 
satellites, the Burst and Transient Source Experiment (BATSE) on board {\it 
Compton Gamma Ray Observatory (CGRO)}, and the currrently available data from 
the Proportional Counter Array (PCA) and High-Energy Transient X-ray Experiment 
(HEXTE) on board \xte.

\section{The distance to \source}
\label{distance}

The distance, $d$, to a source is, obviously, of major importance for 
interpretation of its observations. Therefore, we update here the constraints 
obtained by Z98. They obtained $d\ga 3$ kpc based on the extinction, which, 
however, does not impose an upper limit. In particular, there is a star at 
$\sim$8 kpc within $1\degr$ of \source\ with less extinction than it (fig.\ 
1 of Z98). On the other hand, Z98 found a kinematic distance of $d\simeq 4\pm 
1$ kpc, but they noted that this result was less secure than the extinction 
limit due to a possible peculiar velocity of the system (see below). 

Based on VLT observations of \source, Shahbaz et al.\ (2001) estimated the 
relative contribution of the secondary to the Gunn $r$-band magnitude as $\la 
0.3$.  Thus, the secondary must have an $r$-band magnitude $>$21.4 (note an 
error in Shahbaz et al., who quoted a value of 20.4 on p.\ L19), which 
corresponds, after correcting for $E(B\!-\!V)=1.2$ (Z98), to a late K-type star 
with an apparent diameter of $s \la 3.2\, \rm \mu as$, where we assumed the 
effective temperature of $\sim 4000\, \rm K$. This estimate is also consistent 
with the lowest recorded near-IR magnitudes (work in preparation). In 
particular, $H\simeq 16.4$ observed in 2003 Jan.--Feb.\ (Buxton \& Bailyn 2004), 
combined with the upper limit on the $R$ magnitude of Shahbaz et al.\ (2001) is 
consistent with an early M-type star with $s\sim 3.6\, \rm \mu as$.

Since the system is a LMXB and the secondary is filling its Roche lobe, the mass ratio, $q\equiv M_2/M_\mathrm{X}$, is given by (Eggleton 1983),
\begin{equation}
R_2/a  = \frac{0.49 q^{2/3}}{0.6 q^{2/3} + \ln (1+q^{1/3})},
\label{r_a}
\end{equation}
where $R_2$ is the secondary's radius and $a$ is the separation. Adopting then the orbital parameters of Hynes et al.\ (2003), namely the secondary semi-amplitude of $K_2=317\pm 10\, \mathrm{km\,s}^{-1}$, the period $P=1.7557$ d, and $q\la 0.08$, we obtain $a \sin i \approx 11.9\, \mathrm R_{\sun}$, $R_2/a \sim 0.194$, and $R_2 \sin i \sim 2.3\, \mathrm R_{\sun}$, where $i$ is the inclination. The distance is then $d\simeq 2R_2/s$, which is $\ga 6.7 \sin^{-1}i$ kpc for $s \la 3.2\, \rm \mu as$.

The system is not eclipsing, so $i \la 80\degr$. On the other hand, the 
relatively large secondary mass function and the presence of moderate 
orbital modulation of the $V$ light during its optically-high state (work in 
preparation), is inconsistent with a very low $i$. Assuming then a 
plausible lower limit of $i\ga 45\degr$, we obtain then an inclination-dependent 
lower limit on $d$ of 
\begin{equation}
6.7 \la d_{\rm min}(i) \la 9.4\, {\rm kpc}.
\label{d_min} 
\end{equation}

This result is consistent with the recent lower limit of $d \ga 6$ kpc for the kinematic distance of Hynes et al.\ (2004) based on analysis of high-resolution optical spectra of the Na D lines. Specifically, they find at least 9 line components arising from different clouds located along the line of sight. They derive a range of mean cloud velocity of $-143$ to $+32$ km s$^{-1}$, which they attribute to the Galactic rotation, and argue that \source\ lies at or beyond the tangent point, i.e., $d\ga 6$ kpc. They also suggest that the slightly positive systemic velocity of \source\ and the presence of the absorption component in Na D lines with positive velocity, 25--35 km s$^{-1}$, indicates a location close to the Solar circle on the far side of the Galaxy, i.e., at $\ga 15$ kpc, which distance they favour. Although they note that for such a large $d$, \source\ would be at a $z\sim 1$ kpc below the Galactic plane, they still associate \source\ and the redshifted absorption with the warped material of the outer Galactic disc.

However, we prefer the location in the Galactic bulge for the following reasons. 
(i) The analysis of high resolution spectra Na D and Ca K lines towards OB stars 
(Sembach et al.\ 1993) revealed similar redshifted components in most of their 
sample stars located in the Galactic bulge, at $d\sim 8\pm 2$ kpc, $b\ga 
-\!10\degr$. (ii) The maximum Galactic disc warping on the southern hemisphere 
corresponds to the azimuthal angle of $\Theta = 260\degr$, where the warping 
starts at the radial distance from the Galactic Center of $R=12$ kpc and the 
H{\sc i} disc is bending down to $z=-1$ kpc at $R=16$ kpc (Nakanishi \& Sofue 
2003). Since both $\Theta$ and $R$ are the Galactocentric coordinates, GX 339--4 
cannot be associated with this warped material because at a distance of $d\sim 
15$ kpc, its $\Theta \sim 320\degr$ and $R\sim 8$ kpc. Moreover, $\Theta$ 
increases with the increasing $d$, and a larger $d$ would locate GX 339--4 even 
farther away from the Galactic plane. (iii) The rather high peculiar velocity 
($\sim$140 km s$^{-1}$ if \source\ lies at the tangent point, Hynes et al.\ 
2004) is not unusual if \source\ is a member of the spheroidal Galactic bulge 
population.

Thus, we hereafter use $d=8$ kpc, compatible with the location in the Galactic 
bulge. Using equation (\ref{d_min}), this is also compatible with $i\sim 60\degr$ and $M\sim 10\msun$, given the mass function of $5.8\pm 0.5\msun$ of Hynes et al.\ (2003). We note that the relatively strong Compton reflection measured in X-ray spectra of \source\ (e.g., Z98; W02; Corongiu et al.\ 2003) appears incompatible with $i\simeq 80\degr$; however, an investigation of this issue is beyond the scope of this paper. 

\section{The X-ray data}
\label{data}

The \ginga/ASM (Tsunemi et al.\ 1989) monitored \source\ in the 1--20 keV band from 1987 March 6 to 1991 October 2, and the countrate lightcurve was presented in Kong et al.\ (2002). Here we use instead energy fluxes in the bands of 1--1.6, 1.6--3.3, 3.3--5.0, 5.0--12.5, and 12.5--19.8 keV converted from the instrumental counts using the response matrix of Tsunemi et al.\ (1989).

Using eq.\ (1) of Z02, we measure the hardness across two adjacent energy bands 
in those and subsequent data by a photon spectral index, $\Gamma$, corresponding 
to a power law that would yield the same hardness ratio as observed. Even if the 
actual spectrum is different from a power law, this $\Gamma$ still gives a 
measure of the overall hardness across the two energy bands that avoids the 
dependence on their relative sizes. We do not correct for the interstellar 
absorption ($N_{\rm H}\simeq 6\times 10^{21}$ cm$^{-2}$, Z98). This results in 
the values of $\Gamma$ at $\la$3 keV being somewhat lower than the intrinsic 
ones. 

The used \xte/ASM (Levine et al.\ 1996) data cover the interval of 1996 
January 6--2004 January 29. During that time, \source\ was observed by the ASM 
$\sim$10 times a day on average with each (dwell) exposure of 90 s. Here, we use 
the data binned over one day (from xte.mit.edu/ASM). The daily averages are 
based on dwell data selected using a number of criteria (see that web page), in 
particular the $\chi^2_\nu<1.5$ of the solution. Here, we have imposed an 
additional criterion of the minimum of 4 dwells per day (derived using the ASM 
Crab data), to avoid possible spurious fluctuations. The ASM data have a 
positive bias of 1 mCrab (heasarc.gsfc.nasa.gov), which we have removed based on 
the average ASM Crab counts. 

We then convert the instrumental counts into energy fluxes in the ranges of 
1.5--3, 3--5, and 5--12 keV using the conversion matrix given by eq.\ (A1) of 
Z02, derived based on Cyg X-1 data, but applicable to \source\ as the conversion is independent of the direction. For this work, we have additionally confirmed its correctness using Crab data. We have compared the ASM Crab data with the \sax/MECS data (from www.asdc.asi.it) acquired on 1997 October 8. We have found that the ASM conversion gives the energy fluxes in the above bands $\sim 15$ per cent higher than the MECS data. This is a discrepancy smaller than typical ones between various X-ray instruments (e.g.\ 30 per cent for HEXTE vs.\ PCA). Then, the obtained 3--12 keV effective spectral index for the ASM data is in the 2.04--2.10 range for 200-day averages, in good agreement with $\Gamma=2.09$ for the MECS data. On the other hand, the MECS data give the effective (absorbed) $\Gamma(1.5$--5 keV$)\simeq 1.94$, somewhat softer than the range obtained for ASM of 1.79--1.97. The low relative level of this discrepancy can be illustrated by the fact that it can be completely removed if we modify the lower energy bound from 1.5 keV to 1.55 keV. In addition, the ASM indices above 3 keV of \source\ are in excellent agreement with those from the PCA, see below. 

The BATSE instrument aboard {\it CGRO\/} monitored the X$\gamma$-ray sky 
1991 May 31--2000 May 25. It used Earth occultation to identify point 
sources (e.g., Harmon et al.\ 2002). On average, 12--15 occultations of \source\ 
were observed daily. We use the data binned on a one-day timescale, and reject 
days with $<4$ occultations. We use four energy bins, 20--40, 40--70, 70--160 
and 160--430 keV. The counts have been converted to energy fluxes based on the 
Crab observations of Jung (1989), with the resulting conversion coefficients 
from countrate to energy flux for the four bands of  7.567, 5.830, 8.079, 
$6.887\times 10^{-9}$ erg, respectively. 

\begin{figure}
\centerline{\psfig{file=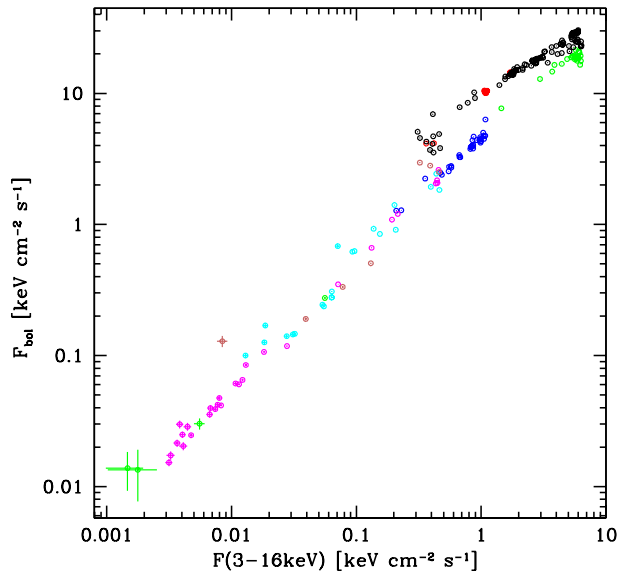,width=8.1cm}} 
\caption{Relation between the intrinsic 3--16 keV flux and the model bolometric flux obtained in fits to the PCA/HEXTE data, see Section \ref{data}. The red and black points correspond to the first and second soft state, respectively, and the blue, magenta, green and cyan points, to the four consecutive hard states before and after those soft states (see Section \ref{lc}). The brown points correspond to the hard state at the beginning of the 2004 outburst. 
}
\label{pca_flux}
\end{figure}

Given that \source\ is a transient source, the data cover wide ranges of fluxes, and often are of limited statistics. We have then rebinned adjacent (i.e., avoiding averaging over large gaps) time bins until achieving a specified accuracy in the 10--30 per cent (10--$3.3\sigma$) range (depending on the considered case), weighing the fluxes with the inverse square error. However, during the off states, achieving those accuracies would require averaging over a very large number of days and often extending the average to include a part of a luminous state. Thus, we also limited the maximum number of averaged days to 100--200. We sum the fluxes and their square errors when adding adjacent energy channels. 

We also analyze all currently available PCA and HEXTE data, following the method 
of Done \& Gierli{\'n}ski (2003, hereafter DG03). We extract one PCA (detectors 
0, 2 and 3, top layer only, except for observations since MJD 52763, when only 
detectors 0 and 2 were available) and HEXTE (cluster 0) spectrum per pointed 
observation, using {\sc ftools} 5.3. We then fit the resulting 236 spectra with 
positive detections by a model consisting of a blackbody disc and thermal 
Comptonization, {\tt thcomp} (Zdziarski, Magdziarz \& Johnson 1996). The effects 
of Compton reflection are approximated by adding smeared edge and Gaussian line. 
The model is absorbed by $N_{\rm H} = 6\times10^{21}$ cm$^{-2}$. We use the 
resulting unabsorbed (unlike the case of the ASM data) energy fluxes in the 
3--4, 4--6.4, 6.4--9.7 and 9.7--16 keV energy bands (as in DG03), normalized to 
the PCA. As above, we convert the hardness ratios to the average spectral 
indices. This method proved to give robust and fairly model-independent fluxes 
as long as the model gives an adequate spectral fit, with $\chi^2_\nu < 1.5$ 
(DG03). This is the case for most of the data, except for a few soft-state 
spectra (with $\chi^2_\nu \sim 2$--3), where the high-energy tail cannot be 
properly reproduced by thermal Comptonization, an indication of the non-thermal 
emission present in soft states (see e.g.\ Gierli{\'n}ski et al.\ 1999; 
Zdziarski et al.\ 2001). This, however, very weakly affects neither the 3--16 
keV flux nor the bolometric one (see below).

We also use the model bolometric flux, $F$, for these data. The systematic error 
related to the uncertainty in the method of calculating $F$ is not included in 
its quoted error (based only on the countrate error). Fig.\ \ref{pca_flux} shows 
the resulting relation between the 3--16 keV flux (with low systematic errors) 
to the obtained model bolometric flux. We see two different relationships in the 
hard and soft states, related to the distinctly different intrinsic spectra of 
the two states, peaking at $\sim$100 keV and $\sim$1 keV, respectively (see, 
e.g., Zdziarski \& Gierli\'nski 2004 for a recent review). 

Finally, we note that we do not study differences in absolute flux 
calibration of different instruments. This adds a systematic error to the 
relative level of different lightcurves. 

\section{Lightcurves} 
\label{lc}

\begin{figure*} 
\centerline{\psfig{file=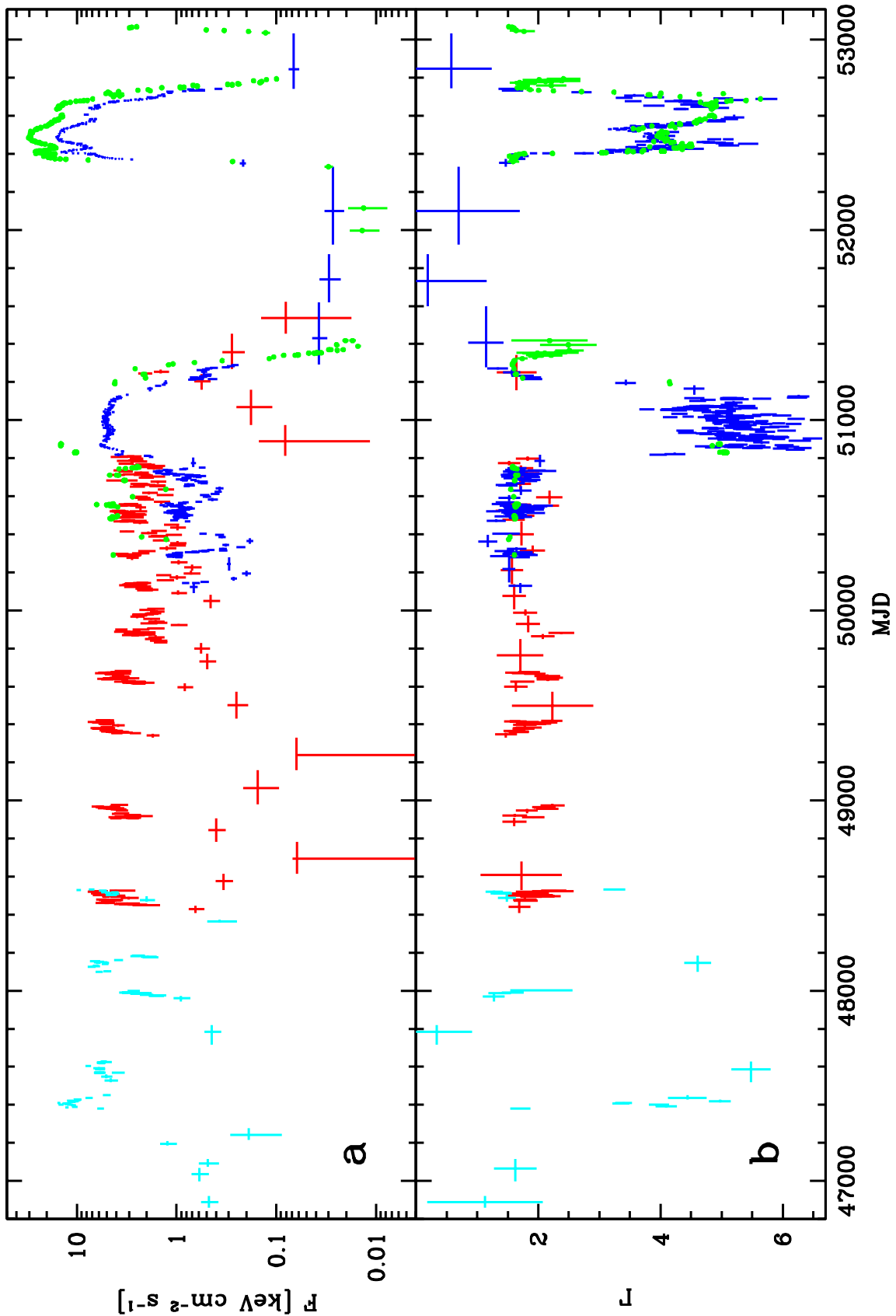,angle=90,width=15.2cm}}
\caption{(a) A long-term lightcurve of \source. The cyan, red, blue and green symbols give the 1--20 keV flux from the \ginga/ASM, 20--430 keV flux from the \gro/BATSE,  1.5--12 keV flux from the \xte/ASM and an estimated bolometric flux from the PCA/HEXTE, respectively. The \xte/ASM points at $<$0.1 keV cm$^{-2}$ s$^{-1}$ are probably overestimated, see Section \ref{lc}. (b) The corresponding average spectral indices in the bands of 3.3--12.5 keV (cyan), 20--70 keV (red), 3--12 keV (blue),  and 3--6.4 keV (green). }
\label{lc_all} 
\end{figure*}

Fig.\ \ref{lc_all}(a) shows the lightcurves based on our data, rebinned to the 
$\ge 5\sigma$ significance. Fig.\ \ref{lc_all}(b) shows the corresponding 
evolution of selected average spectral indices (based on data rebinned to $\ge 
10\sigma$ for clarity of display). The indices of $\Gamma\la 2$ (at energies 
$>$3 keV) correspond to the hard state, and those with $\Gamma\ga 2.2$, to soft 
states. 

The \ginga/ASM first saw some weak hard states, and then a strong soft 
outburst around MJD 47400, classified as including the very high state (VHS) 
based on \ginga/LAC observations (Miyamoto et al.\ 1991). (See Zdziarski \& 
Gierli\'nski 2004 for spectra and discussion of that state.) After the flux 
declined by a factor of at least several, another soft peak (possibly a 
continuation of the first outburst) occured at MJD $\sim$47600. The next 
outburst occured at MJD $\sim$48100, with the hard state followed by the soft 
one. 

\begin{figure*} 
\centerline{\psfig{file=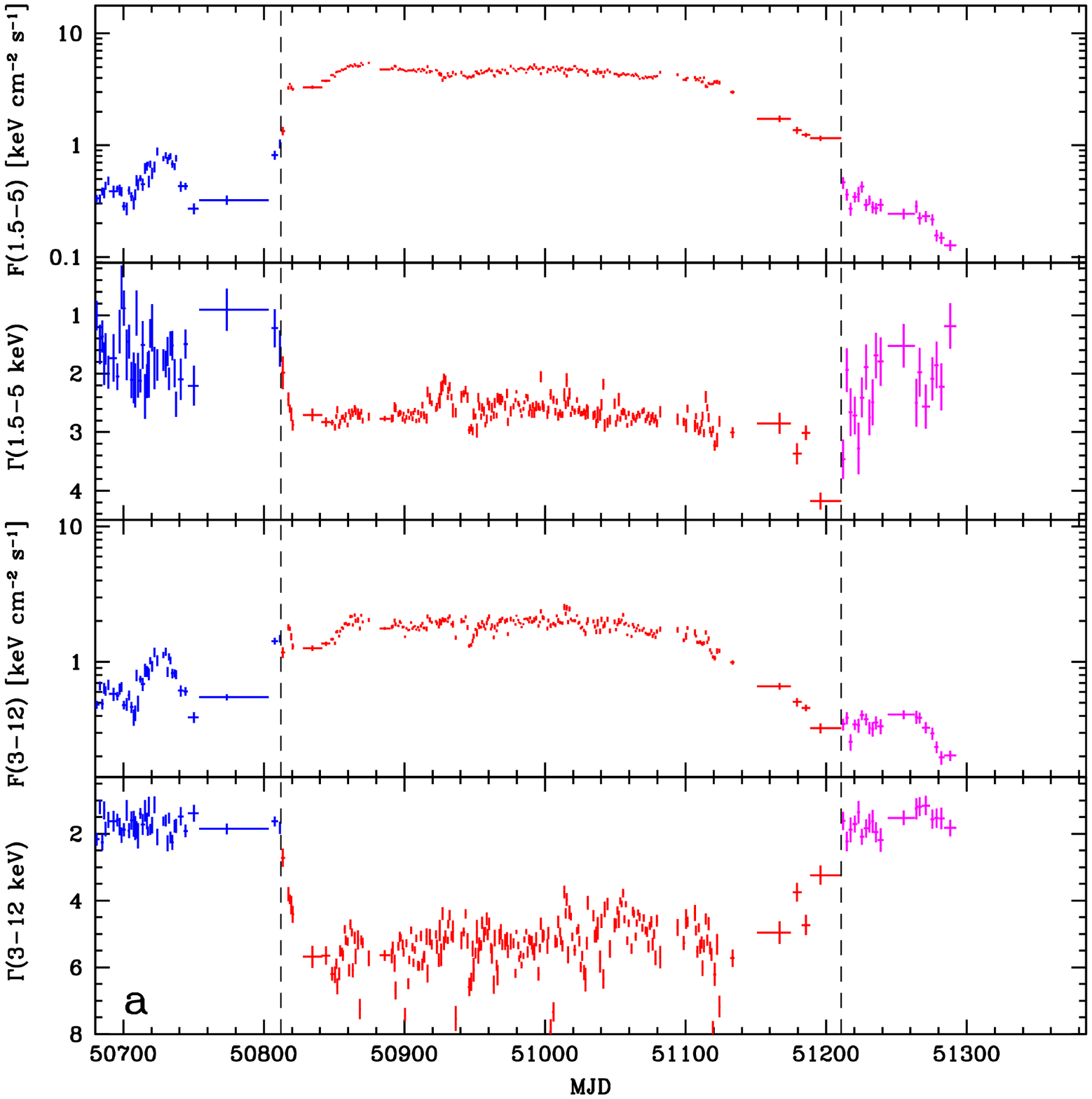,height=11.cm}  \psfig{file=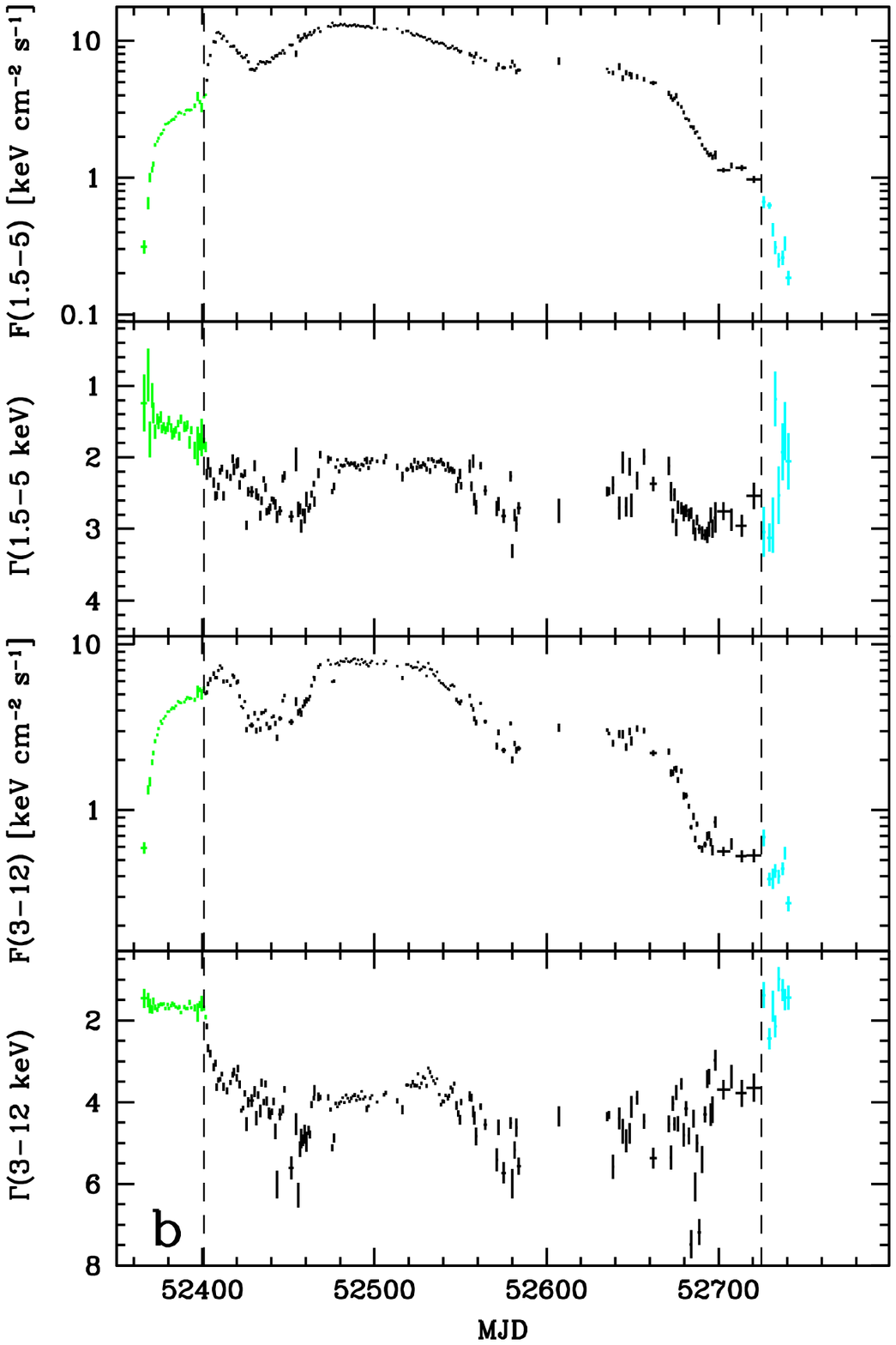,height=11.cm}}
\centerline{\psfig{file=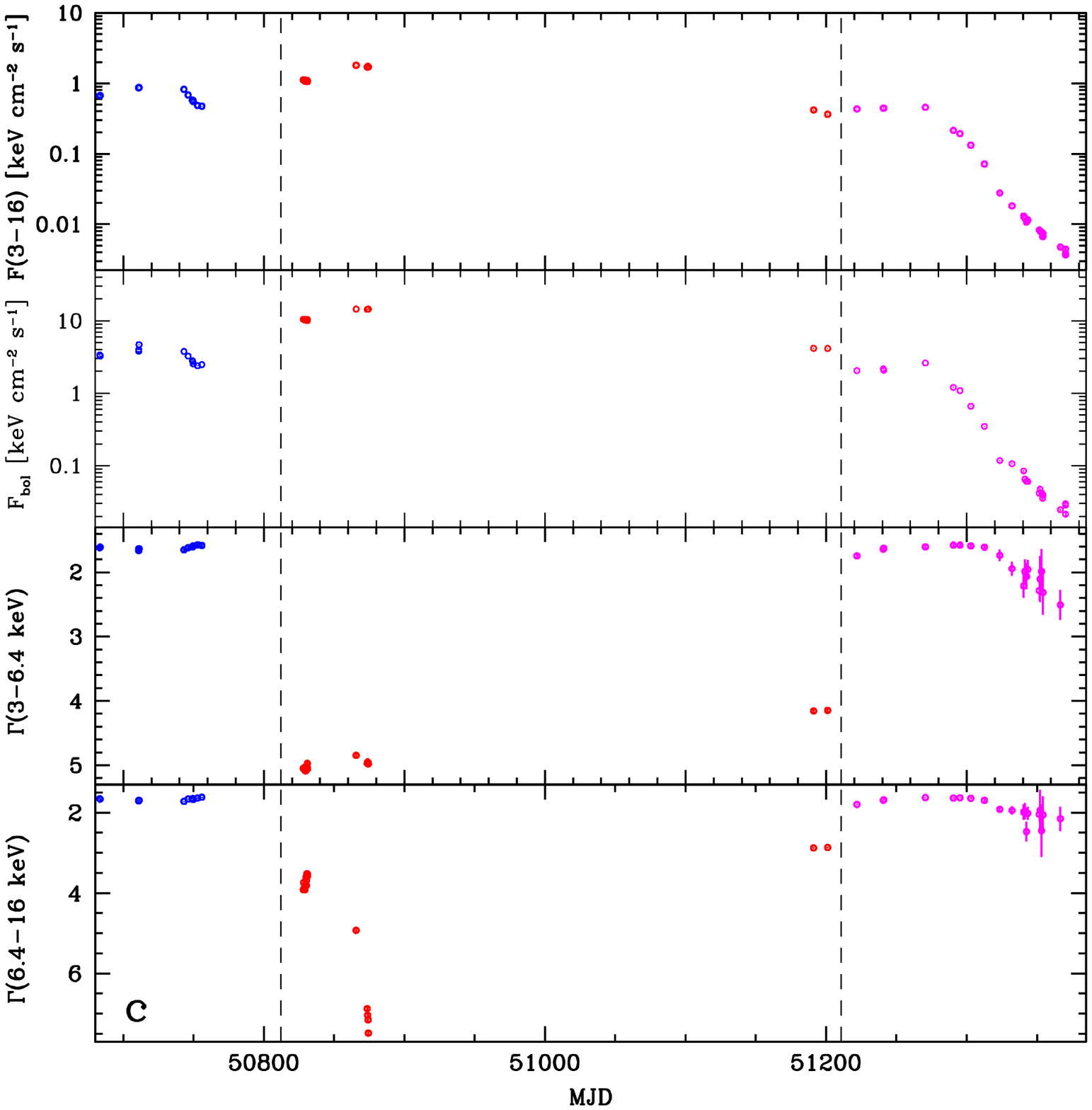,height=11.1cm}  \psfig{file=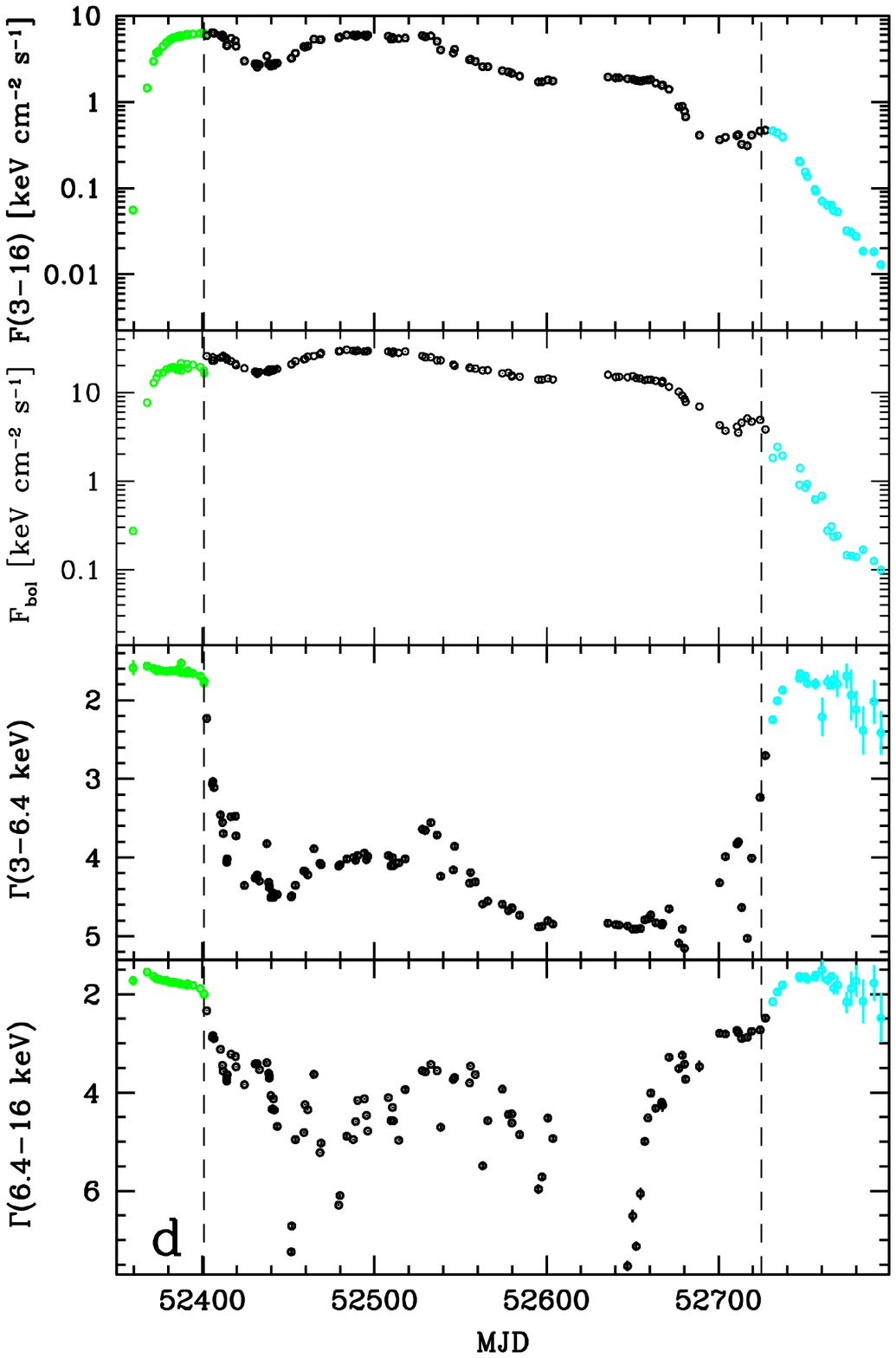,height=11.1cm}}
\caption{The \xte\/ ASM (top; rebinned to $\ge 5\sigma$) and PCA/HEXTE (bottom) fluxes, $F$, and average spectral indices, $\Gamma$, vs.\ time during two outbursts states. Colours mark consecutive phases of the outbursts, with red and black corresponding to the two soft states, and blue/magenta and green/cyan, to the hard state states before/after each of the two soft outbursts. State transitions are marked by vertical lines. The first 600 days (see Fig.\ \ref{lc_all}) of the hard state observed by \xte\/ (blue) are not shown. } 
\label{outbursts} 
\end{figure*}

Another outburst occured at MJD $\sim$48500. The initial hard state seen by the 
BATSE was followed by the soft one, only beginning of which was recorded by 
\ginga. The next two outbursts took place at MJD $\sim$49000 and $\sim$49400. 
The source became very weak between the outbursts, with $F(0.5$--10 keV$) \la 
2\times 10^{-4}$ keV cm$^{-2}$ s$^{-1}$ measured by \asca\/ on MJD 49246 (Asai 
et al.\ 1998). Until MJD 49400, our data had shown 5 major outbursts occuring 
every $\sim$500 d. However, the character of the variability changed thereafter. 
The outbursts became progressively more frequent, weaker, and with the 
increasing interim minimum fluxes (Rubin et al.\ 1998). The intervals 
between the peaks became $\la 200$ d. Also, the \xte/ASM and PCA/HEXTE 
monitoring showed only the hard state until MJD $\sim$50800, see Fig.\ 
\ref{lc_all}.

Then, the variability pattern changed again with \source\ entering an extended 
soft state lasting for $\sim 400$ d. After its decline, a hard state took place, 
and then the source gradually faded. This phase was well covered by the 
PCA/HEXTE, which showed very low fluxes during the resulting off 
state. Also, the \sax\/ observation on MJD 51403 showed $F(2$--10 keV$) \sim 10^{-3}$ keV cm$^{-2}$ s$^{-1}$ (Kong et al.\ 2000). We note there were PCA observations on MJD 51396 and 51418, which both show the 3--16 keV flux $\sim$4 times higher. This may be due to variability, or may indicate some problems with PCA observations of \source\ at such low fluxes. 

On the other hand, rebinning of the \xte/ASM data gave the flux level of $\sim 
0.03$ keV cm$^{-2}$ s$^{-1}$, which may be an overestimate due to, e.g., a 
residual contribution of the Galactic ridge emission (see W02). An indication of 
such a problem is also a discrepancy between the derived ASM hard indices and 
those from the PCA, much softer, see Fig.\ \ref{lc_all}. Thus we hereafter not 
take into account the fluxes obtained by rebinning the \xte/ASM data during off 
states. 

This off state lasted $\sim$1000 d, until MJD $\sim$52360. Then another outburst 
took place (Smith et al.\ 2002a), with the hard-soft-hard spectral variability 
pattern. The outburst ended at MJD $\sim$52800.  Its flux level also has 
exceeded (Zdziarski \& Gierli\'nski 2004) for the first time the highest 
previously recorded level of MJD $\sim$47400. The \source\ went then to a 
quiescent state, with an observation by {\it Chandra\/} showing the unabsorbed 
$F(0.4$--11 keV$) \simeq 2\times 10^{-4}$ keV cm$^{-2}$ s$^{-1}$ and 
$\Gamma\simeq 2.0\pm 0.2$ on MJD 52911 (Gallo et al.\ 2004). This is identical 
to the lowest previous upper limit from \asca\/ (see above). 

The newest outburst was detected on MJD 53036 (Smith et al.\ 2004; 
Belloni et al.\ 2004; Kuulkers et al.\ 2004), as shown in Fig.\ \ref{lc_all} 
till MJD 53068. The initial PCA data show the source in its standard hard 
state, with $\Gamma\sim 1.5$. 

Of particular interest are the two recent complete outbursts covered by the 
\xte/ASM and PCA/HEXTE, showing the hard-soft-hard spectral evolution. Fig.\ 
\ref{outbursts} shows the ASM and PCA/HEXTE fluxes and spectral indices in 
detail. The two soft outbursts (red and black symbols) are similar in shape, but 
their 1.5--12-keV flux and duration are higher and lower by factors of $\sim$2 
and $\sim$1.5, respectively, in the second outburst. A likely cause of this 
difference is the behaviour before the soft state. Before the first outburst, 
\source\ underwent an extended period of a variable hard state, during which its 
broad-band flux was persistently relatively high for $\sim 10^3$ d, whereas the 
second outburst followed a $\sim 10^3$ d period of the off state, see Fig.\ 
\ref{lc_all}.

The vertical lines marking the state transitions in Fig.\ \ref{outbursts} are at MJD 50812, 51211, 52401 and 52725. The last number is based on the ASM data, which show a strong hardening on that day, see Fig.\ \ref{outbursts}(b). On the other hand, the PCA/HEXTE data favour a slightly later soft-to-hard transition day, MJD 52731, consistent with a softening of the 3--12 keV ASM spectrum after the hardening on MJD 52725. 

Apart from the different maximum fluxes in the two outbursts, another remarkable 
property is the varying flux level corresponding to transitions between the 
states. The hard-to-soft transition of the second outburst occured at the level 
higher by a factor of $\sim$3 (similar to the ratio between the maximum 1.5--12 
keV fluxes of the two outbursts) than that of the first one. In both cases, this 
level was higher than that of the soft-to-hard transition, which happened at the 
very similar flux levels for both outbutsts. 

The lagging of the effect in a physical system when the acting force is changed 
is called hysteresis. In the present case, the acting force is apparently the 
accretion rate and the effect is the spectral state. The presence of hysteresis 
in \source\ has been found before by, e.g., Miyamoto et al.\ (1995), Smith 
et al.\ (2002b), Nowak, Wilms \& Dove (2002), and by, e.g., Maccarone \& Coppi 
(2003) and Rossi et al.\ (2004) in some other black-hole and neutron-star binaries. 

We see in Fig.\ \ref{outbursts} that the 3--12 and 3--16 keV fluxes change only 
slightly but the 1.5--5 keV one changes by a large factor during each state 
transition. This implies a transition pivot at $\sim$10 keV, similarly to Cyg X-1 (Zhang et al.\ 1997). 

Another interesting effect is seen during the soft-to-hard transitions. Namely, 
the hardening of the 3--12 keV spectrum is accompanied by an initial softening 
of the 1.5--5 keV one during the first outburst, see Figs.\ \ref{outbursts}(a). 
Such softening is also seen in the 3--6.4 keV spectrum near the end of the 
second soft state, see Fig.\ \ref{outbursts}(d). This effect likely corresponds 
to disappearance of inner parts of the optically-thick disc when its luminosity 
declines at the end of a soft state. Then, at the beginning of a hard state, a 
strong soft excess is still present in the spectrum.

\section{Correlations between the spectral indices and fluxes}
\label{colours}

\subsection{Low energies}
\label{lowE}

\begin{figure*}
\centerline{\psfig{file=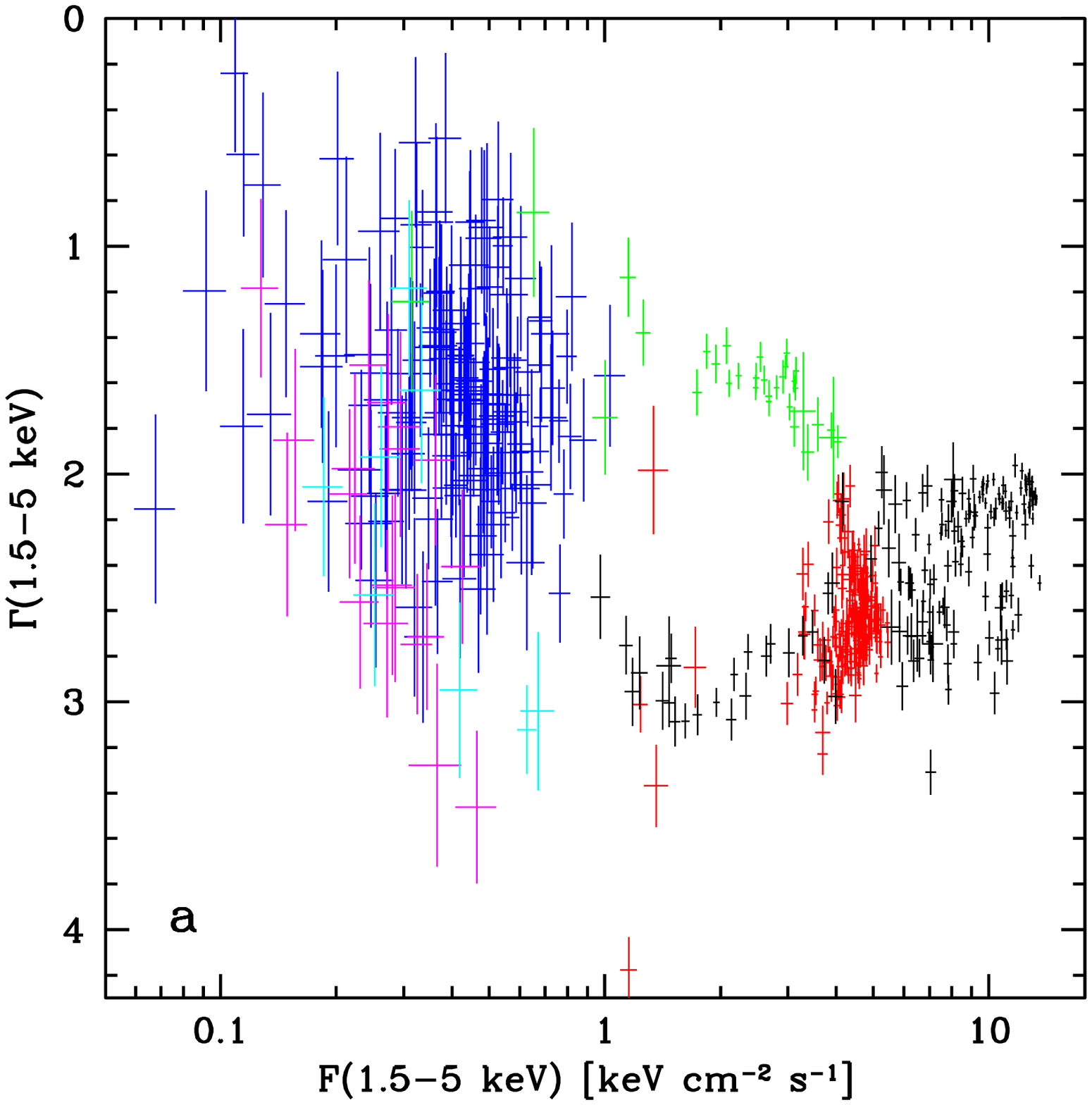,width=7.cm} \psfig{file=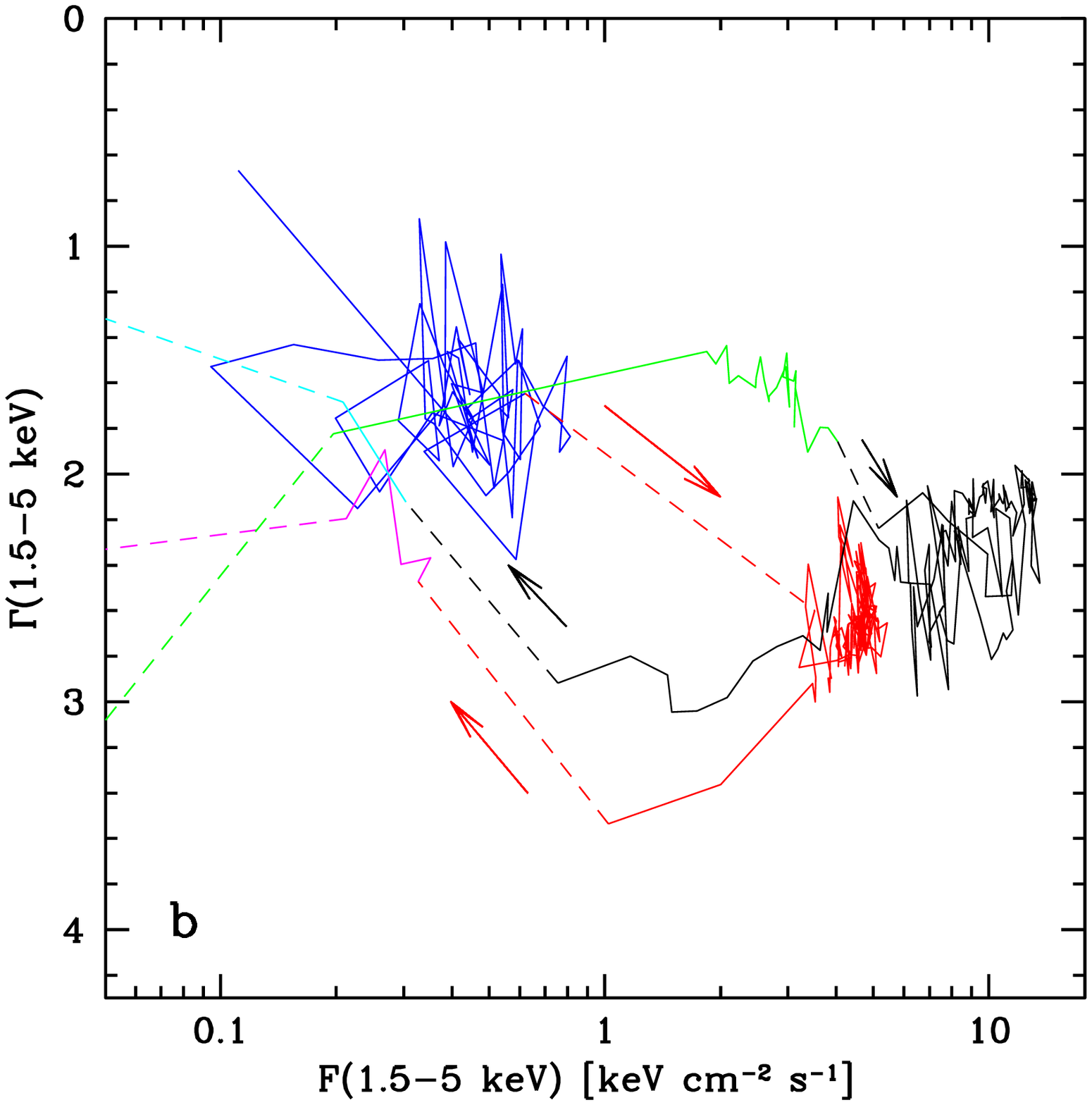,width=7.cm}}
\centerline{\psfig{file=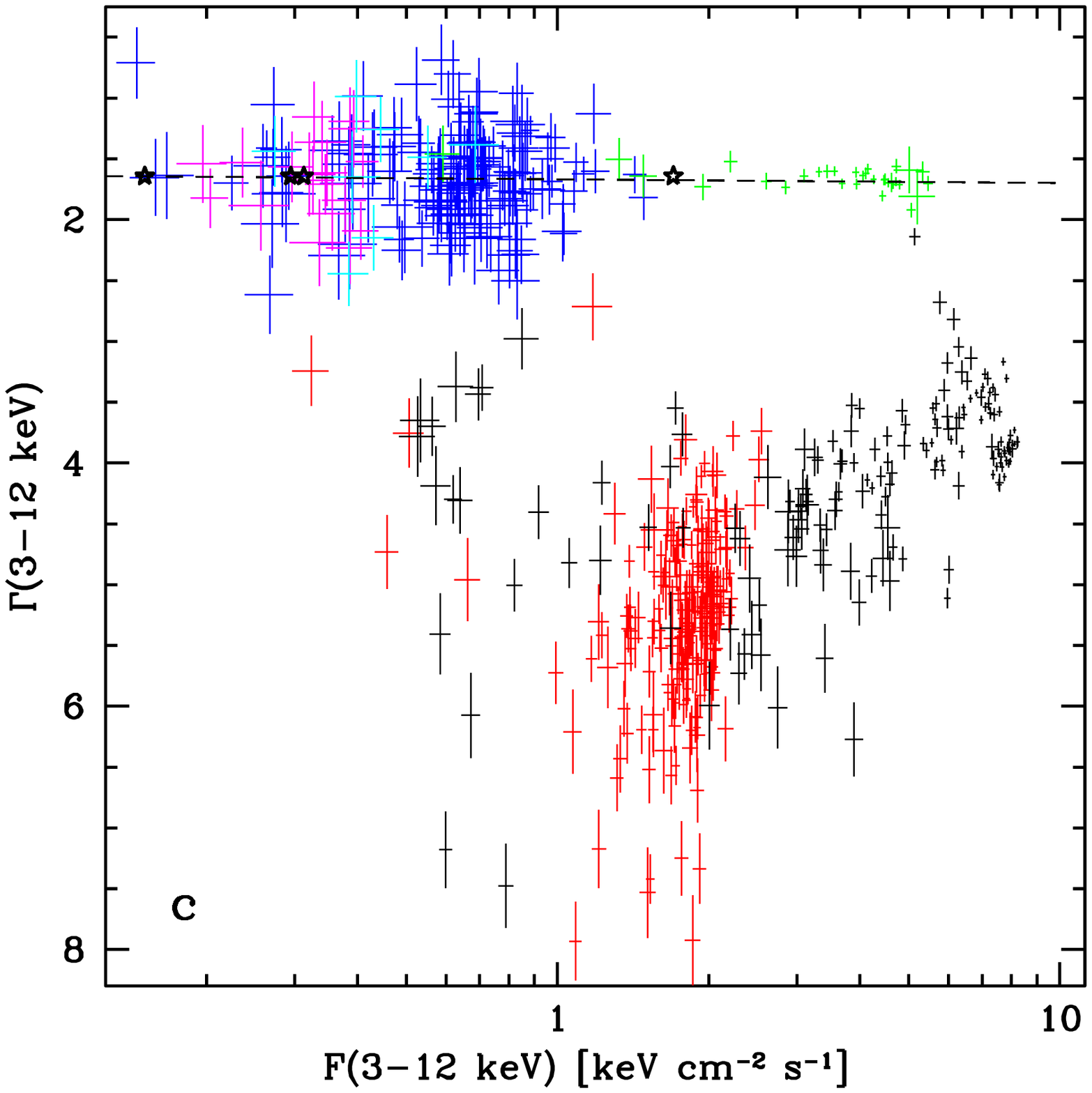,width=7.cm} \psfig{file=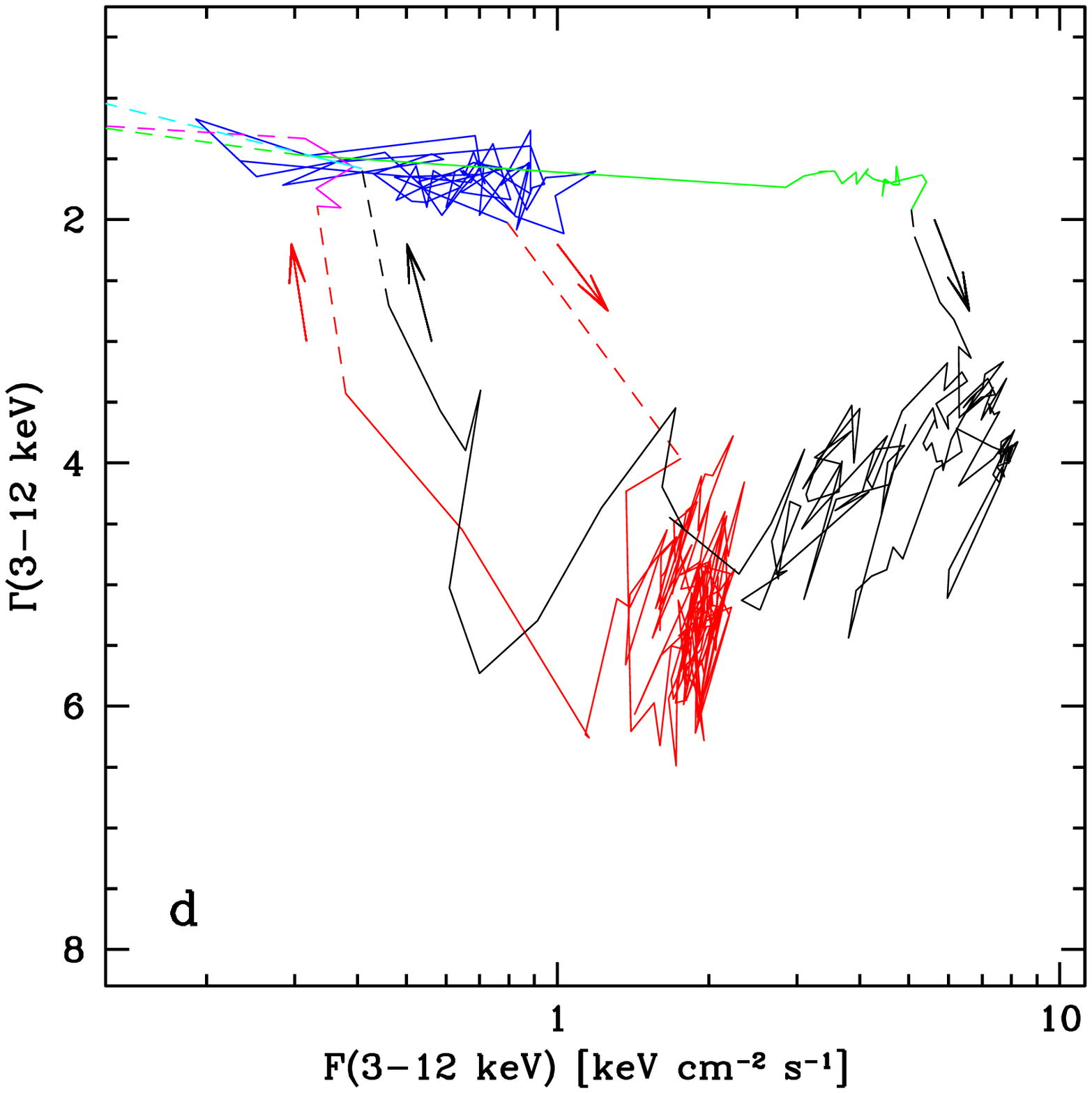,width=7.cm}}
\centerline{\psfig{file=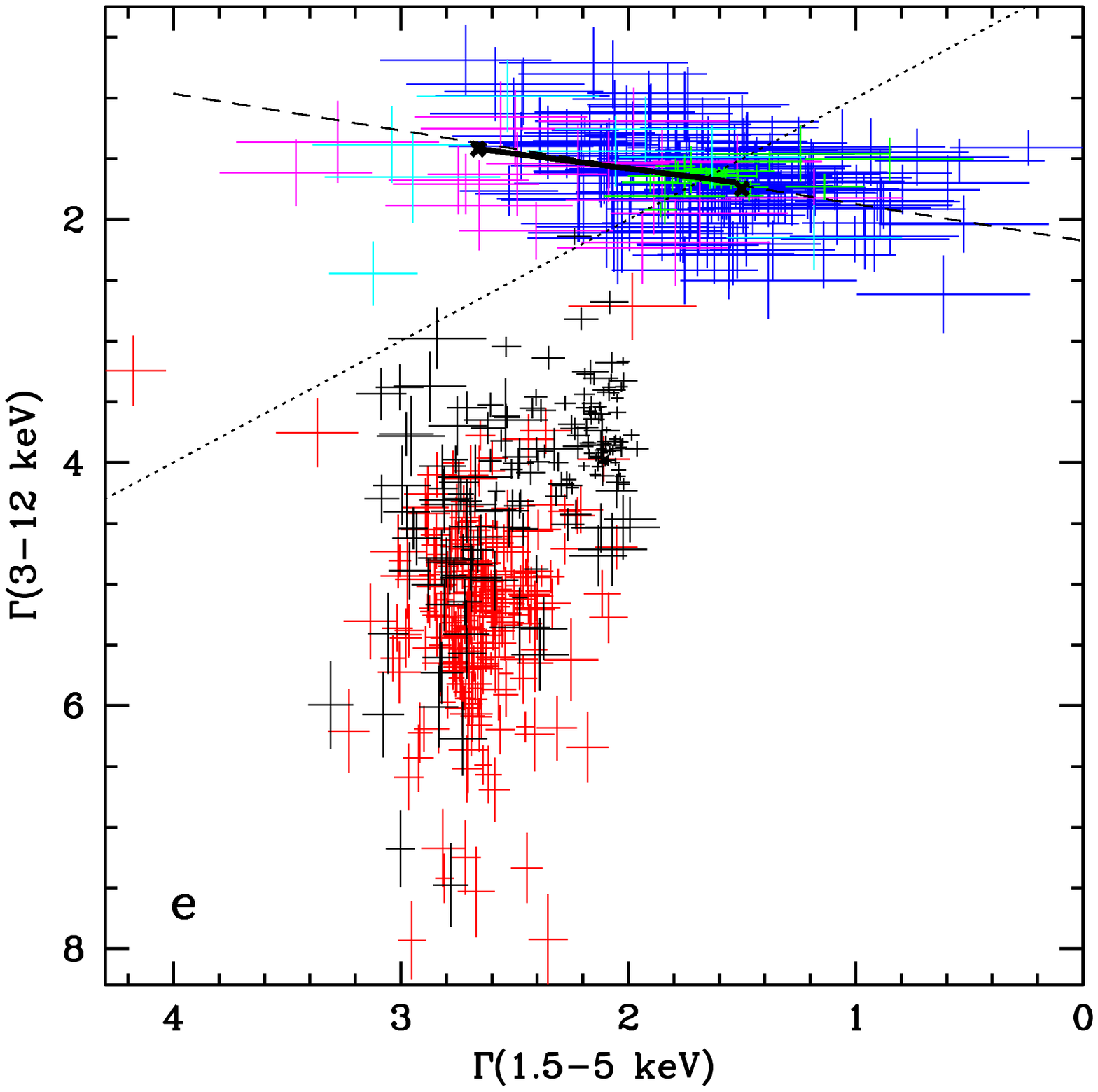,width=7.cm} \psfig{file=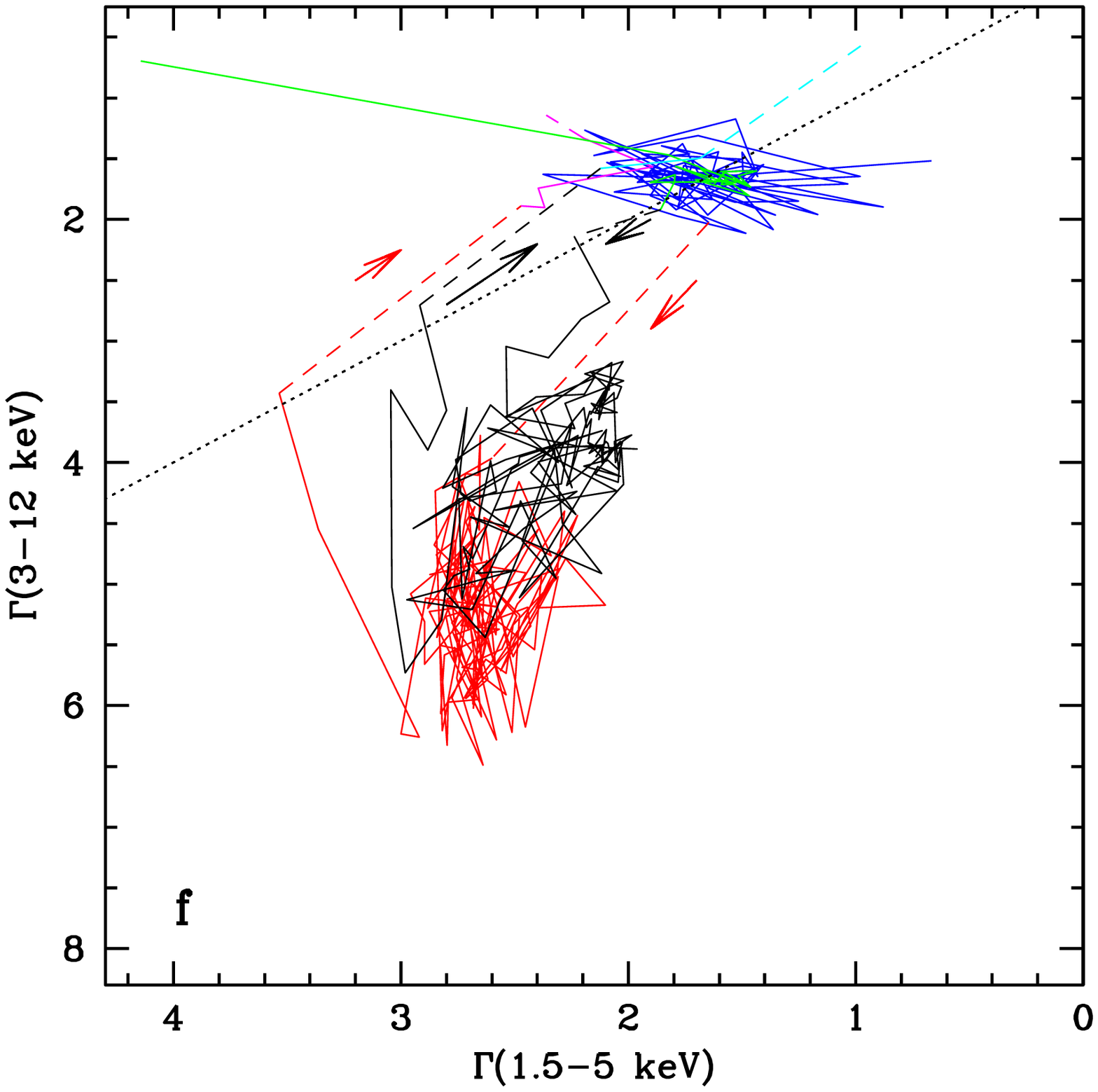,width=7.cm}}
\caption{Relationships between the \xte/ASM spectral indices and fluxes in the 1.5--5 keV and 3--12 keV energy bands. Colours correspond to the outburst phases shown in Fig.\ \ref{outbursts}. The black asterisks in (c) correspond to the \ginga/OSSE pointed observation (the rightmost symbol) and models for decreasing $L$, see Section \ref{hard}. The dashed lines in (c), (e) correspond to the best linear fits to the data, and the dotted lines in (e), (f) correspond to the two indices being equal. The heavy solid line in (e) corresponds to a model with a variable soft photon flux incident on a hot plasma, see Section \ref{hard}. In the right panels, the error bars are suppresssed and lines connect the points in the temporal order (with arrows showing the direction of time). Dashed lines connect points of differrent states; in particular, some of the dashed lines in (b) and (d) show transitions to/from the off state. The data have been rebinned to the $\ge 5\sigma$ and $10\sigma$ significance in the left and right panels, respectively. }
\label{asm_index}
\end{figure*}

Fig.\ \ref{asm_index} presents correlations in the ASM data, including those 
shown in Fig.\ \ref{outbursts}(a--b). We plot the indices decreasing along the 
axes to keep a correspondence to hardness ratio diagrams. In the 1.5--5 keV 
range, the $F$-$\Gamma$ region occupied by the data forms roughly a circle, see 
Fig.\ \ref{asm_index}(a). Evolution is clockwise, see Fig.\ \ref{asm_index}(b). 
The second outburst is extended to fluxes higher than the first. Interestingly, 
both the hard and soft state (defined based on the 3--12 keV spectrum) occupy 
similar ranges of the 1.5--5 keV spectral indices and differ mostly in their 
fluxes. 

This situation reverses in the 3--12 keV range, see Figs.\ \ref{asm_index}(c), 
(d). The hard state occupies an almost horizontal strip for $\Gamma\simeq 
1.5$--2 clearly separated from the soft state but the range of $F$ is similar in 
both states, only slightly shifted to higher values for the soft state. In fact, 
the first soft state (red symbols) shows fluxes typically lower than the hard 
state preceding the second soft state. There is at most slight $\Gamma(\log_{10} 
F)$ correlation in the hard state, with the best fit slope of $0.03\pm 0.02$. We 
note, however, that our $\Gamma(3\!-\! 12\,{\rm keV})$ does not separate the 
Compton reflection component, which strength positively correlates with the 
intrinsic slope (Ueda, Ebisawa \& Done 1994; Revnivtsev, Gilfanov \& Churazov 
2001). Thus, the range of variability of the intrinsic $\Gamma$ will be slightly 
larger than that shown in Fig.\ \ref{asm_index}(c). Fig.\ \ref{asm_index}(d) 
shows clearly the presence of hysteresis, with the transitions to the soft state 
occuring at much larger fluxes than those of out that state. A part of the 
horizontal extent covered in each state is due to this effect, in which the 
spectrum remains relatively stable within a large range of the flux. We also see 
a separation between the regions corresponding to the two outbursts, consistent 
with the difference between their flux scale, Fig.\ \ref{outbursts}.

In both 1.5--5 and 3--12 keV bands, the spectral index is anticorrelated with 
the flux in the soft states, and the anticorrelation is much stronger during the 
second outburst (black symbols). This effect is probably caused by the spectral 
shape varying due to the variable amplitude of the high energy tail beyond the 
blackbody component (as it is the case in Cyg X-1, Z02). For many points, 
especially during the first outburst, the 3--12 keV power law is very soft, up 
to $\Gamma\sim 7$, i.e., the emission in the tail becomes very weak. 

Figs.\ \ref{asm_index}(e), (f) show the corresponding colour-colour diagram. The 
two indices are positively correlated in the soft state, which simply means that 
softening at lower energies is accompanied by softening at higher energies. The 
dotted line in Fig.\ \ref{asm_index}(e) corresponds to the two indices being 
equal. Thus, we see that in the soft state the 3--12 keV spectrum is in most 
cases softer than the 1.5--5 keV one. This softening at high energies probably 
corresponds to both the cutoff in the disc blackbody entering the range of 
energies $>$3 keV and absorption.

Our strongest hard-state (anti)correlation involves the 1.5--5 and 3--12 keV spectral indices, see Fig.\ \ref{asm_index}(e). We see that the harder the 3--12 keV spectrum, the softer the 1.5--5 keV one, i.e., the stronger the soft X-ray excess. The slope of this correlation is $-0.30\pm 0.05$ (obtained by fitting including errors in both variables, e.g., Press et al.\ 1992), and the Spearman correlation coefficient is $-0.37$, corresponding to the probability of occuring by chance of $\sim 10^{-8}$. The part of the hard-state data to the right of the dotted line appears to correspond to the hardening of the lower-energy spectrum due to absorption. On the other hand, there is a soft X-ray excess to the left of the dotted line, strong enough to be seen in spite of the absorption. As seen in Fig.\ \ref{asm_index}(f), the left-most part of the hard state corresponds to the return from the soft state, and the soft excess here is probably a residual from that state. Still, similar soft excesses are also clearly present in the hard-state data (blue symbols) before the first soft state. We discuss a theoretical explanation of this correlation in Section \ref{hard}.

Overall, this diagram strongly resembles those for atoll neutron-star binaries, e.g., Aql X-1 (fig.\ 4 in DG03), but {\it not\/} the colour-colour diagrams for black-hole binaries, which are usually quite diagonal (fig.\ 2 in DG03), which is also the case for the corresponding ASM plot for Cyg X-1 (fig.\ 5 in Z02). 

\begin{figure}
\centerline{\psfig{file=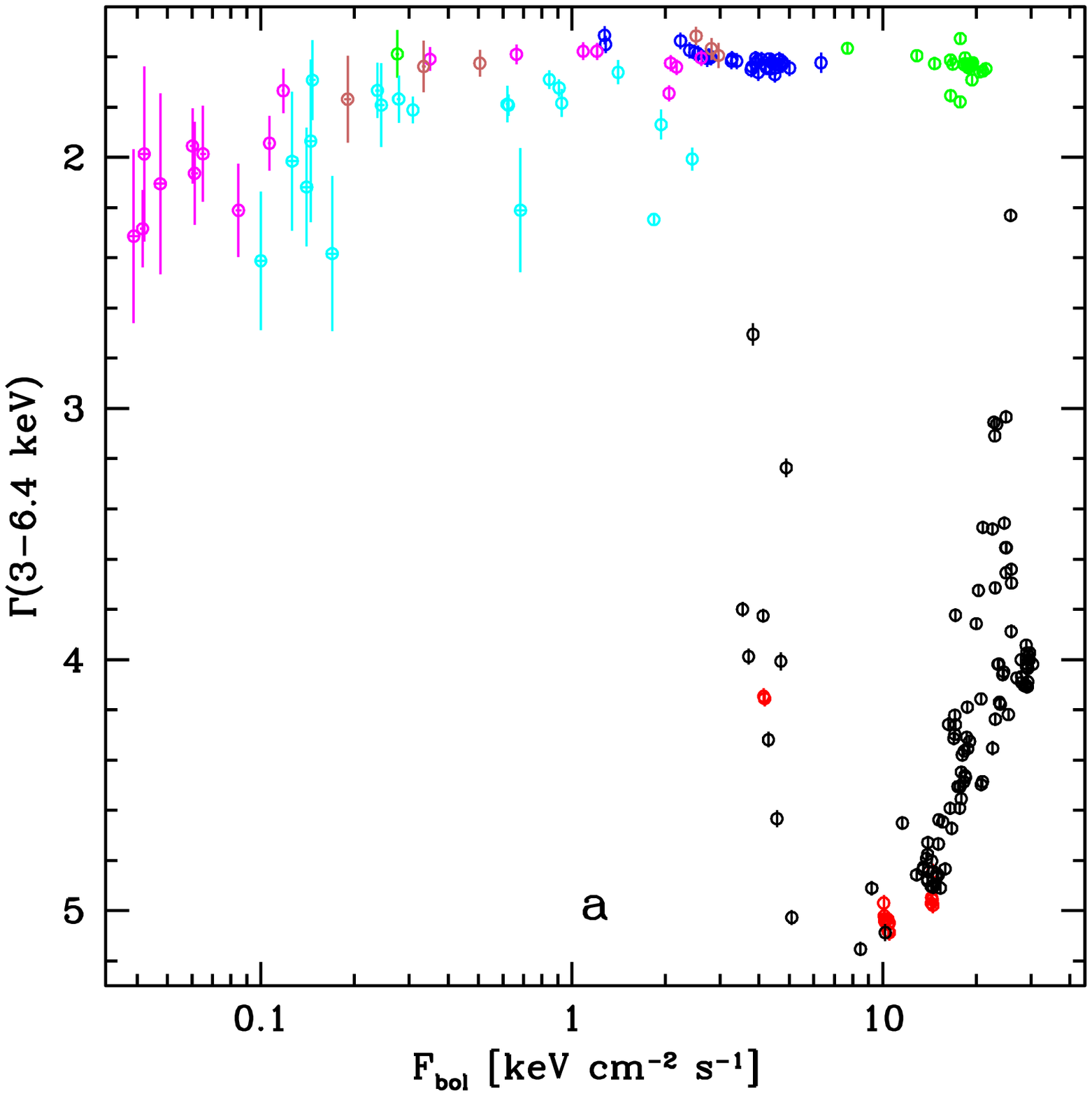,width=7.cm}} \centerline{\psfig{file=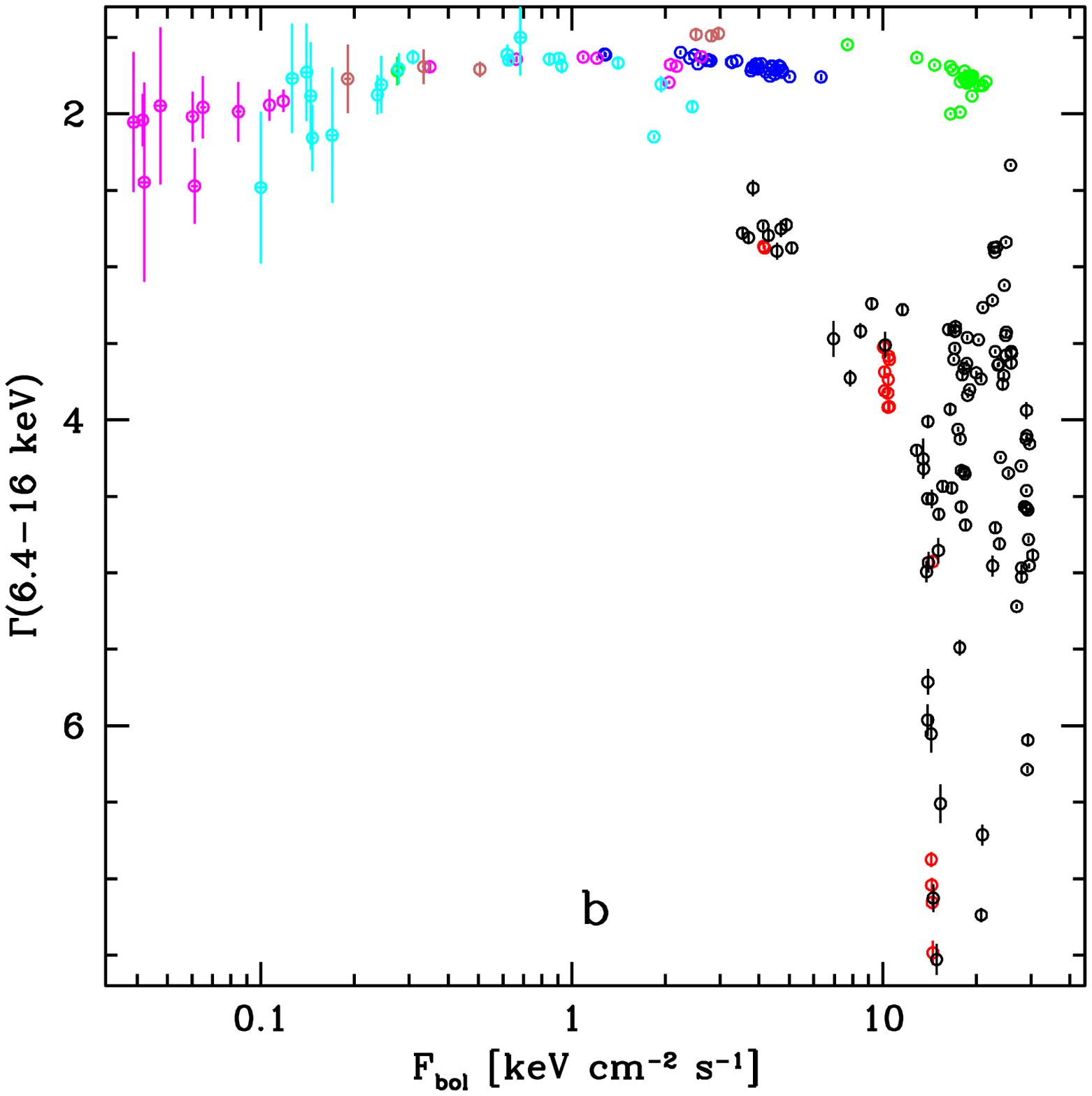,width=7cm}}
\centerline{\psfig{file=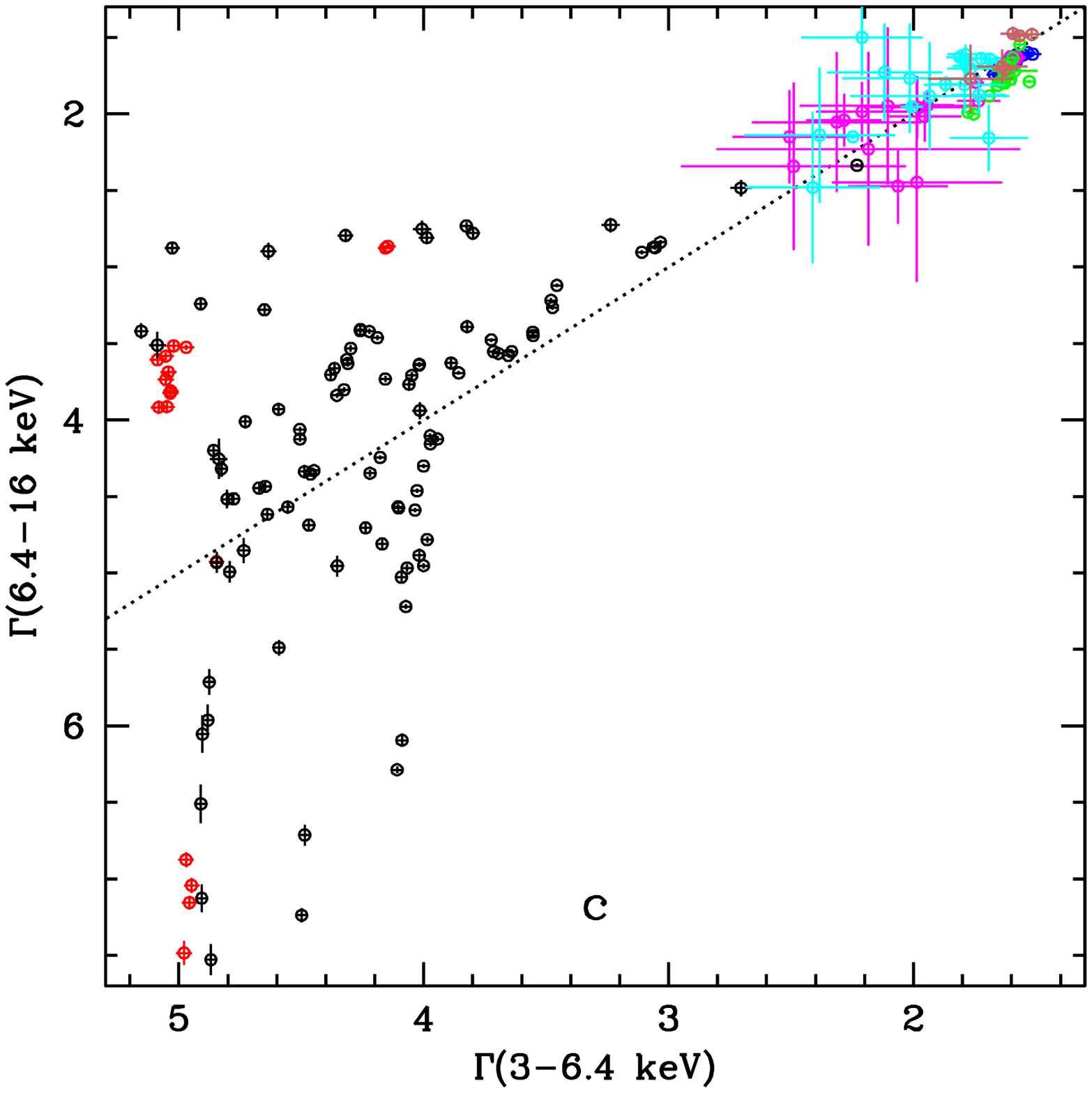,width=7.cm}}
\caption{Relationships between the 3--6.4 keV and 6.4--16 keV \xte/PCA/HEXTE spectral indices and estimated bolometric fluxes. The dotted line in (c) corresponds to the two indices being equal. The colours have the same meaning as in Figs.\ \ref{outbursts}, \ref{asm_index} except for brown corresponding to the hard state at the beginning of the 2004 outburst. The outburst evolution in (a) and (b) is clockwise.}
\label{pca_index}
\end{figure}

However, an important difference between colour-colour diagrams based on the 
ASM and PCA (used by DG03) data is the energy range. The ASM covers energy 
$\ga$1.5 keV, whereas the coverage of the PCA is for $\ga$3 keV. In particular, 
the colours used by DG03 are for 3--6.4 and 6.4--16 keV, i.e., at energies above 
the lowest ASM channel, 1.5--3 keV. The hard-state soft excess in \source\ is 
clearly constrained to energies $\la$3 keV. Fig.\ \ref{pca_index} shows diagrams 
corresponding to those in Fig.\ \ref{asm_index}(c), (e) but based on the 
PCA/HEXTE, see Fig.\ \ref{outbursts}(c--d). We see in Fig.\ \ref{pca_index}(c) 
that the hard state colour-colour diagram for \source\ is approximately 
diagonal, like for other black-hole binaries (DG03).

\begin{figure}
\centerline{\psfig{file=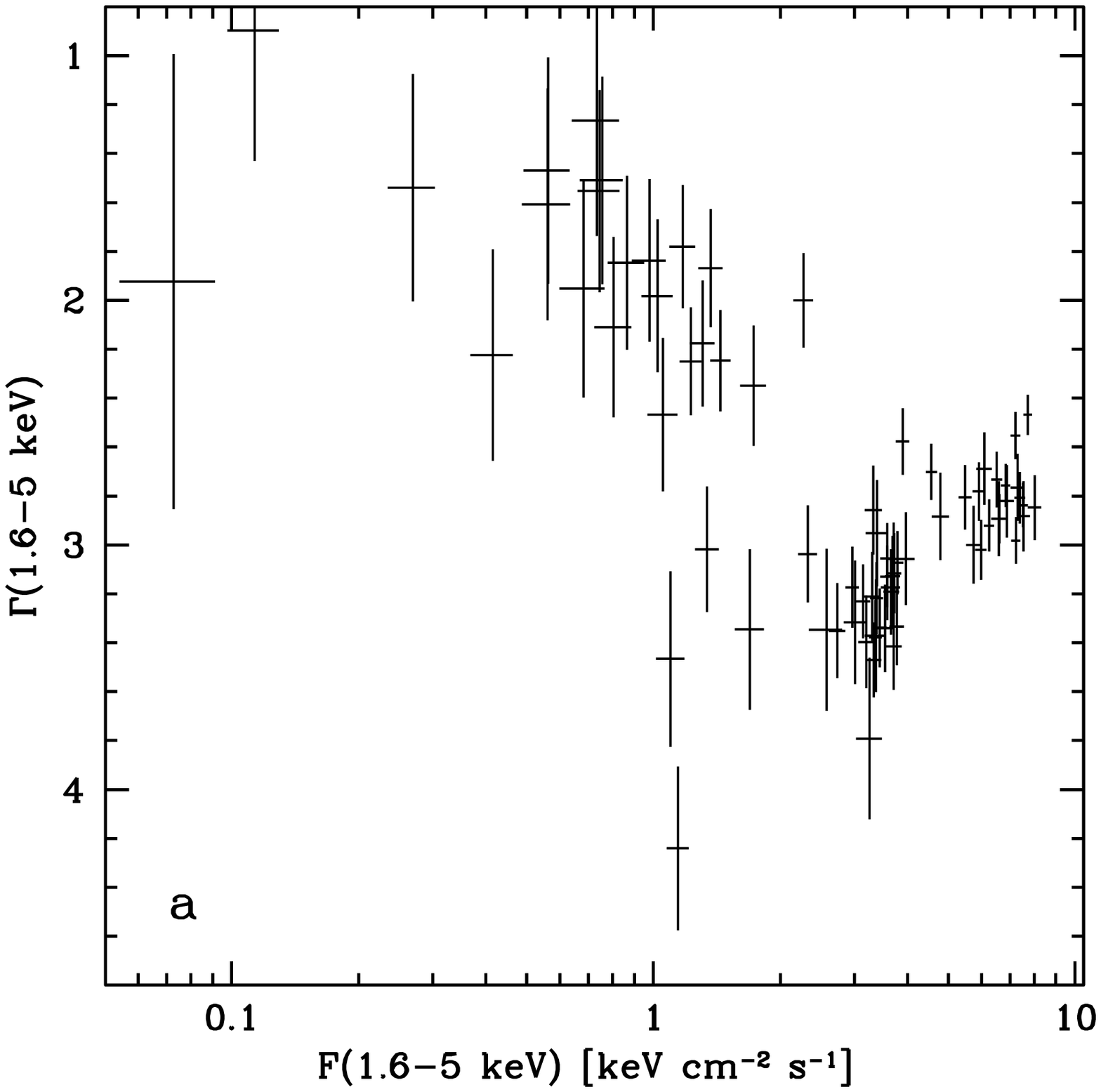,width=4.2cm} \psfig{file=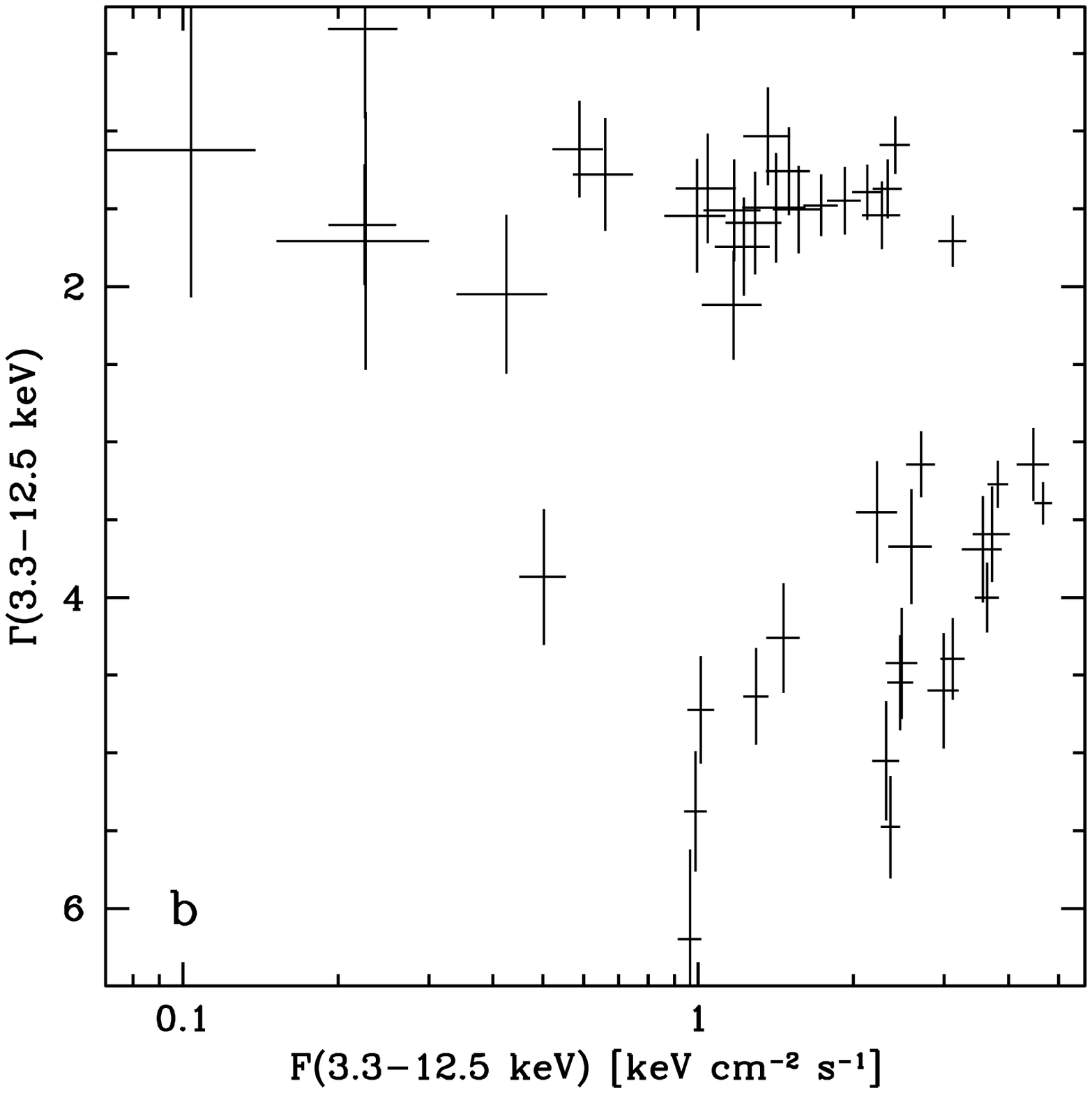,width=4.2cm}}
\caption{Relationships between the 1987--91 \ginga/ASM spectral indices and fluxes in the 1.6--5 keV and 3.3--12.5 keV energy bands (rebinned to the $\geq 5\sigma$ significance). We see good agreement with the 1996--2003 \xte/ASM results on Figs.\ \ref{asm_index}(a), (c). }
\label{ginga_index}
\end{figure}

Given that \source\ is definitely a black-hole system (Hynes et al.\ 2003), the difference between the hard state of atoll and black-hole binaries in their colour-colour diagrams at energies $\ga$3 keV (with the former and latter forming horizontal and diagonal regions) cannot be attributed solely to the cooling action of the neutron star surface, which was proposed by DG03. Instead, the difference is, at least partly, related to the different blackbody temperature, $T_{\rm bb}$, in the two classes of sources. The upper horizontal branch appears due to the disc blackbody emission affecting the colours, which obviously depends on the maximum $T_{\rm bb}$. This temperature is significantly higher in neutron-star systems due to their lower masses, and thus the effect of blackbody emission is seen above 3 keV, whereas it is not seen in black-hole systems with a lower temperature. However, the upper horizontal branch, related to the hard-state blackbody, still appears in our black-hole system when low enough energies are studied. Thus, black-hole and neutron-star binaries are more similar than the results DG03 would suggest.

Another interesting property is that whereas Fig.\ \ref{asm_index}(c) shows a softening from the lower to the higher energies for virtually all soft-state data, the soft state in the corresponding Fig.\ \ref{pca_index}(c) shows either hardening or softening from energies below 6.4 keV to those above it. This difference is also related to the different energy ranges in the two diagrams. In the ASM range, we see mostly the effect of the high-energy cutoff of the disc blackbody, whereas in the PCA range, we see the variable high-energy tail (which can be quite hard) beyond this cutoff. 

Figs.\ \ref{pca_index}(a), (b) show the two PCA indices as a function of the 
model bolometric flux. In the hard state, both indices are almost constant at 
high fluxes, and appear to soften at low fluxes ($\la$0.1 keV cm$^{-2}$ s$^{-1}$). Although the soft state extends to the highest fluxes, we see many soft-state points corresponding to the bolometric fluxes {\it below\/} the maximum fluxe of the hard state. This clearly shows the presence of the hysteresis, now based on the broad-band PCA/HEXTE data. Fig.\ \ref{pca_index} includes the data for the beginning of the 2004 outburst, showing it to be in the standard hard state. 

\begin{figure*}
\centerline{\psfig{file=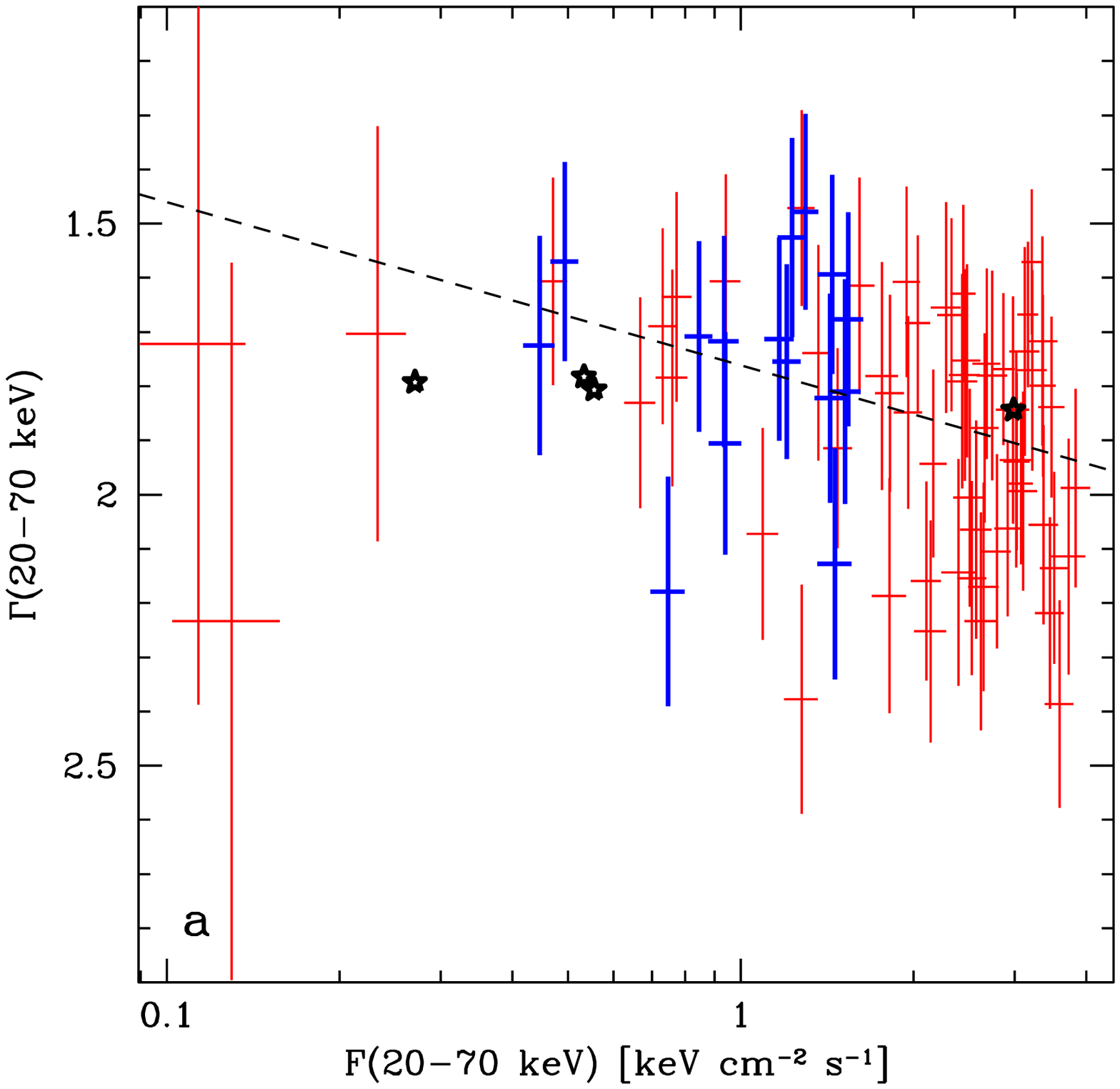,width=6.cm}  \psfig{file=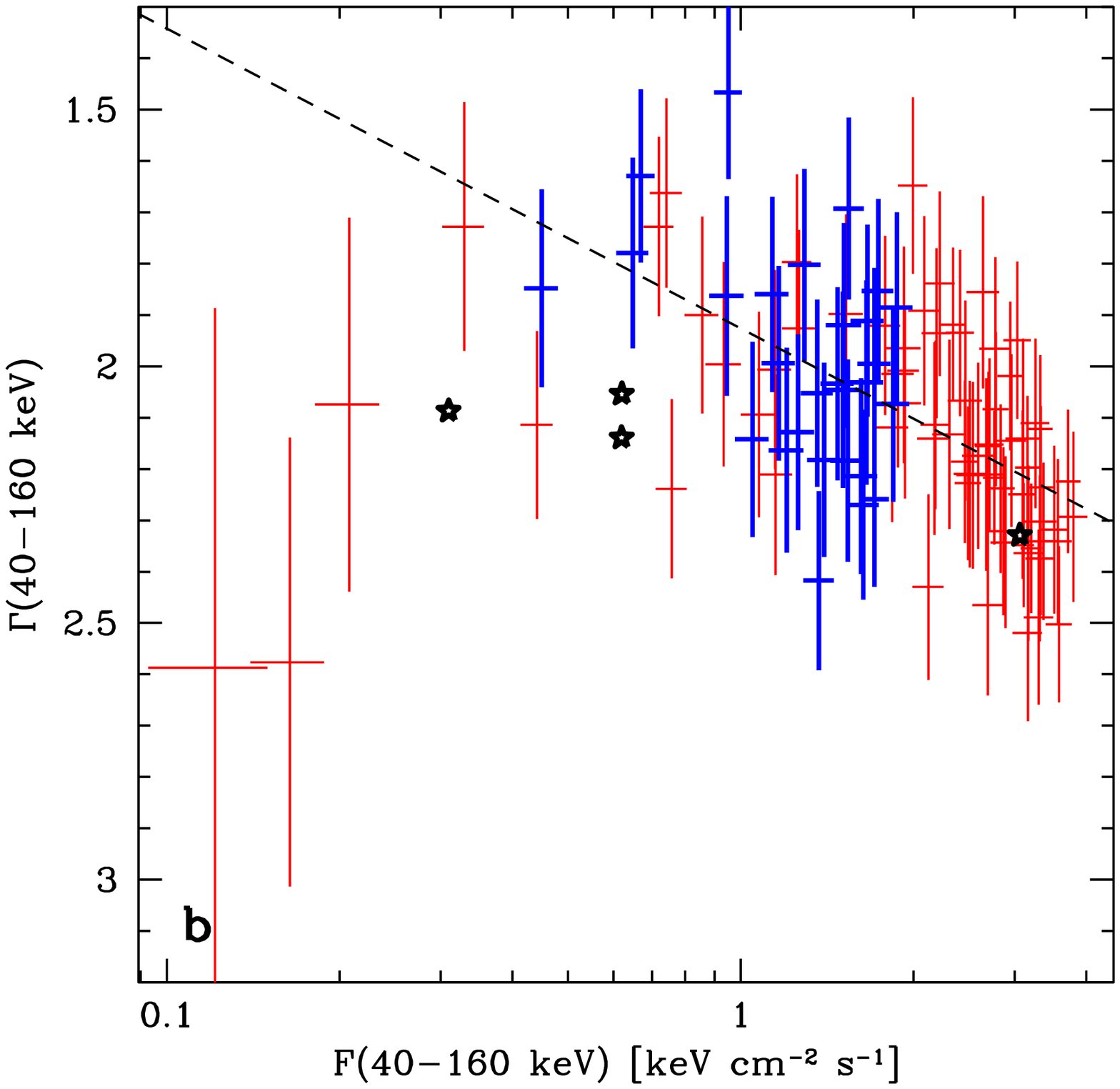,width=6.cm}}
\centerline{\psfig{file=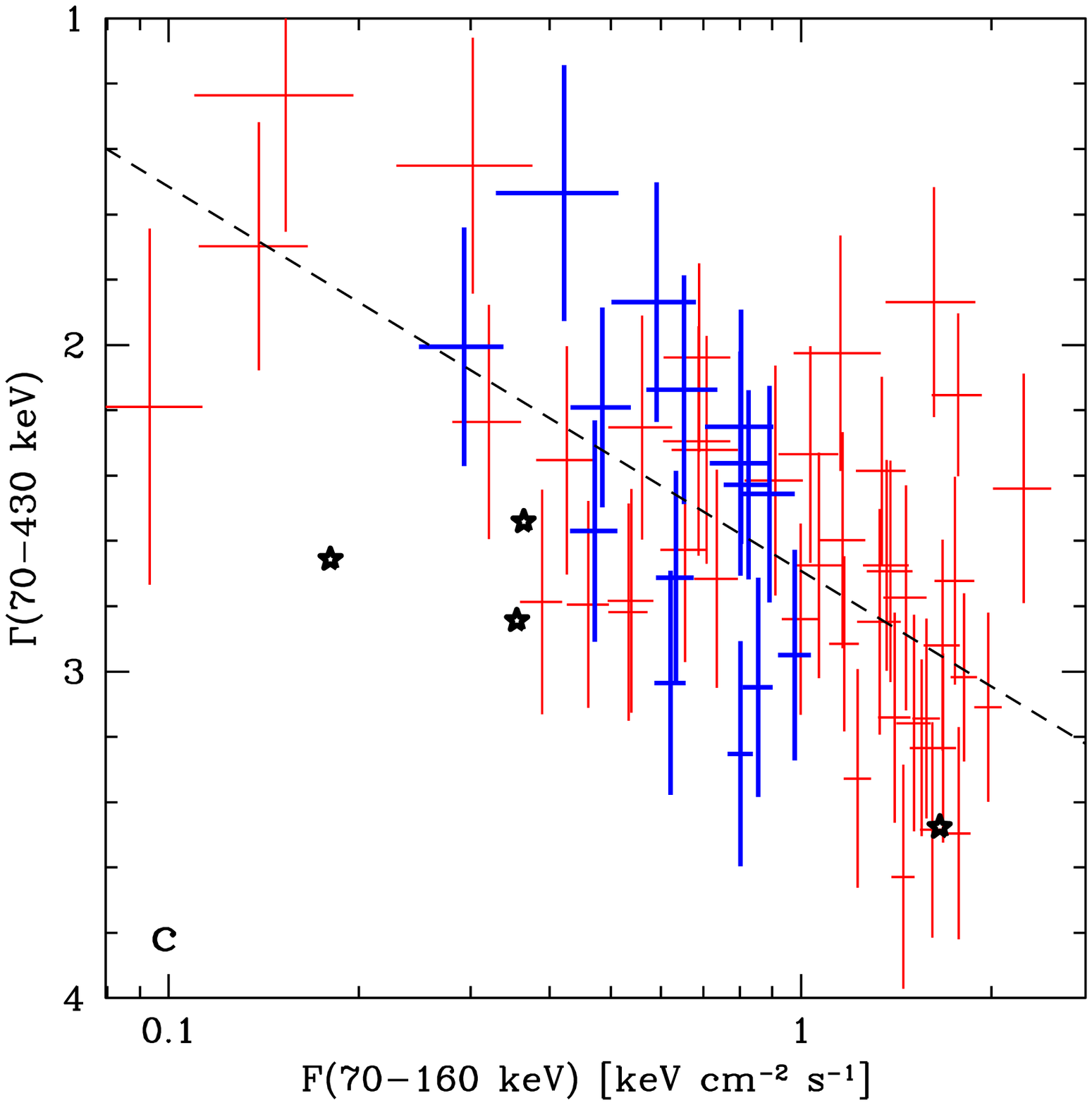,width=6.cm} \psfig{file=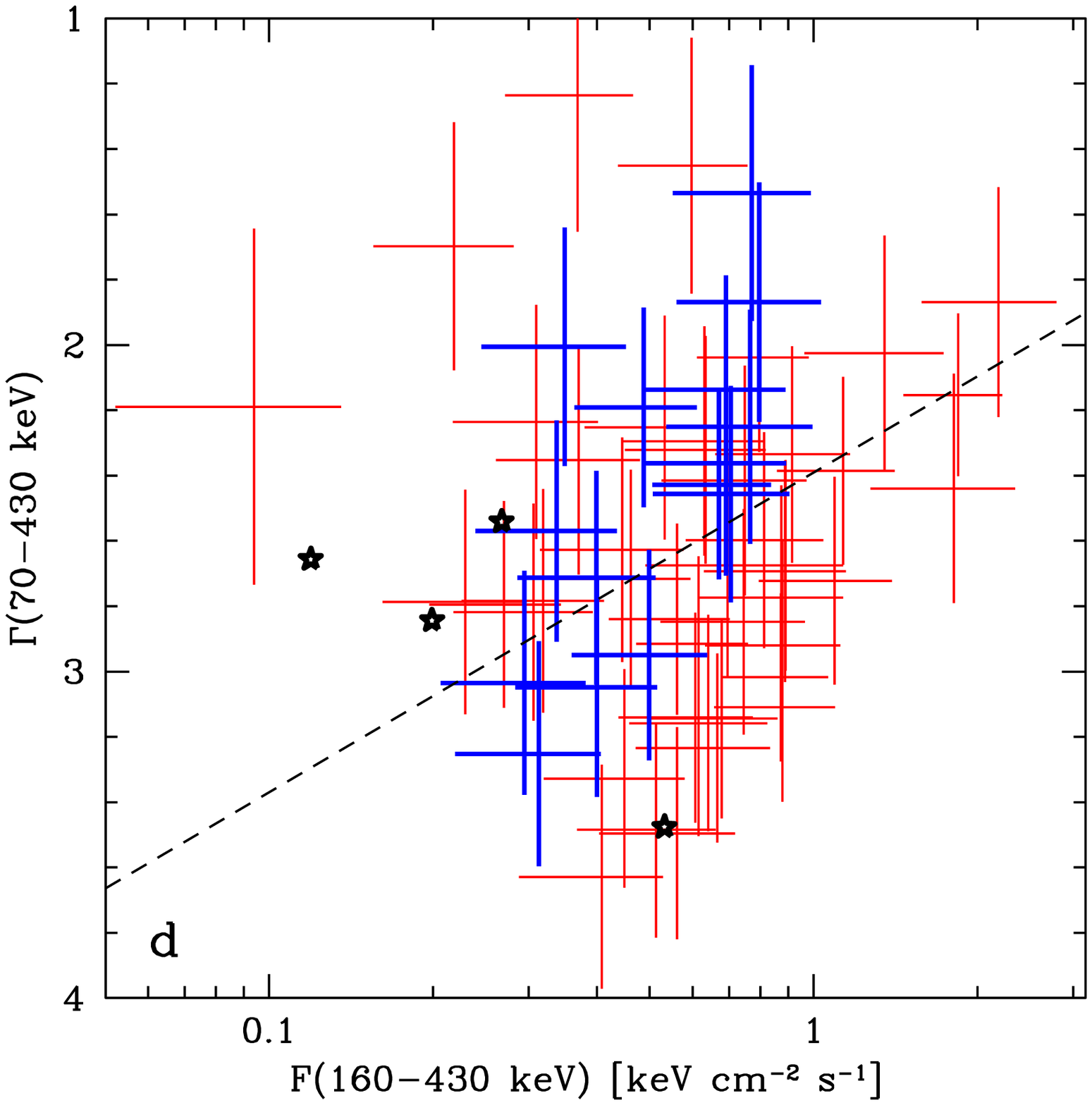,width=6.cm}}
\caption{Relationships between \gro/BATSE indices and fluxes in the 20--430-keV energy bands. The heavy blue crosses correspond to the hard state covered by the \xte/ASM (Figs.\ \ref{outbursts}, \ref{asm_index}). The thin red ones correspond to the preceding BATSE observations, which were in the hard state except for the leftmost points. The data for (a--b) and (c--d) have been rebinned to the $10\sigma$ and $3.3\sigma$ significance, respectively. The dashed lines show the best linear fits. The asterisks have the same meaning as in Fig.\ \ref{asm_index}, see Section \ref{hard}. }
\label{batse_index}
\end{figure*}

Fig.\ \ref{ginga_index} shows the colour-flux diagrams based on the \ginga/ASM analogous for those for the \xte/ASM, Figs.\ \ref{asm_index}(a), (c). We see virtually identical dependences, showing that no significant change has occured in \source\ in about a decade separating those observations. 

\subsection{High energies}
\label{highE}

Fig.\ \ref{batse_index} shows the relationships between the effective spectral 
indices in three bands vs.\ fluxes for the BATSE data in the hard state covered 
by the \xte/ASM (blue) and for earlier epochs (red). Most of the symbols also 
correspond to the hard state, as the soft state is hardly detected by the BATSE. 
Those diagrams represent continuation of those for the \xte/ASM,  Figs.\ 
\ref{asm_index}(a), (c), towards higher energies. 

In the 3--12 keV band, $\Gamma$ was almost independent of $\log_{10} F$, with 
the best fit slope of $0.03\pm 0.02$. Now, we see clear $F$-$\Gamma$ 
correlations from 20 to 160 keV, with its strength increasing with energy. 
Linear $\Gamma(\log_{10} F)$ fits to all the hard-state points give the slopes 
of $0.30\pm 0.09$ and $0.58\pm 0.08$ for the 20--70, 40--160 keV bands, 
respectively, see Figs.\ \ref{batse_index}(a), (b). The corresponding Spearman 
correlation coefficients are 0.39 and 0.62, with the chance probabilities of the 
correlation of $5\times 10^{-4}$ and $10^{-11}$, respectively. Thus, it is very 
unlikely that those correlations are due to a chance occurence, and the strength 
of the correlation given by the slope increases with the photon energy. (Note that the strength of the correlations is due to the BATSE data preceding the ASM monitoring, when the strongest fluxes were recorded.) 

This result means that the softening of the spectrum with increasing flux 
becomes progressively stronger with increasing energy. It indicates the high 
energy cutoff in the spectrum decreasing with the increasing flux, as also found 
by W02 based on broad-band pointed observations. We discuss a theoretical explanation of this result in Section \ref{hard}. 

The dependence of $\Gamma(70\!-\! 430\,{\rm keV})$ on the flux is more complex. 
There is now a change of the sign of the correlation in the middle of the 
70--430 keV energy band, which, as we have checked, does not appear in the 
previous two cases. Namely, there is still very strong positive correlation of 
this index on the 70--160-keV flux, with the slope of $1.18\pm 0.16$, and the 
Spearman coefficient of 0.54 corresponding to the chance probability of 
$1.5\times 10^{-6}$, see Fig.\ \ref{batse_index}(c). The sign of the 
correlation, however, reverses for the dependence on the 160--430 keV flux, as 
seen in Fig.\ \ref{batse_index}(d), with the best-fit slope of $-1.0\pm 0.3$. 
These effects are seen for both the BATSE data simultaneous with the \xte/ASM 
and the earlier ones. This reversal of the correlation sign shows the hard-state 
variability can be described by a variable slope at high energies with the pivot 
point at $\sim$200 keV. 

Note that the pivoting variability also contributes to a part of the softening of the 20--160 keV slopes with increasing flux. However, this softening cannot be solely due to pivoting as its degree increases with the photon energy, opposite to the pivoting, where the $F$-$\Gamma$ correlation becomes stronger in the opposite direction, i.e., away from the pivot energy.  

We then considered the joint ASM/BATSE simultaneous data, dominated by the hard 
state. We have found that any of the three ASM fluxes is well linearly 
correlated with any of the BATSE fluxes in the bands of 20--40, 40--70 and 
70--160 keV energy bands. However, there is no clear correlation of the ASM 
fluxes with the 160--430 keV band. This is consistent with Fig.\ 
\ref{batse_index}(d), showing a reversal of the positive flux-index correlation 
above 160 keV. 

\section{Variability properties}
\label{variability}

\subsection{Fractional variability}

\begin{figure}
\centerline{\psfig{file=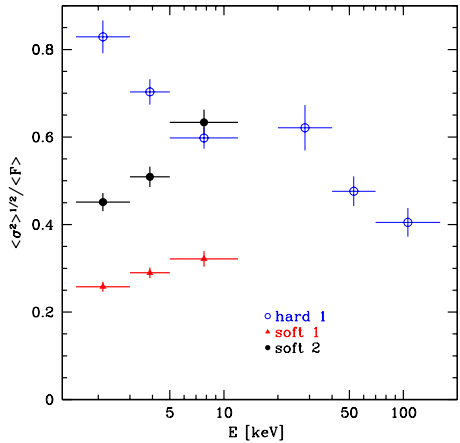,width=6.cm}} 
\caption{The fractional rms variability vs.\ photon energy from the ASM and BATSE data.
}
\label{rms}
\end{figure}

Additional information about the physical states of the source is provided by 
the fractional rms variability as a function of energy. We calculate it 
separately for the hard state covered jointly by the \xte/ASM and BATSE, i.e., 
until the first soft state, and for each of the two soft states using only the 
ASM data. In general, there are several ways to compute the rms, e.g., using 
data rebinned up to a given significance. This, however, depends on the chosen 
significance level. Thus, we have decided to show results obtained using the raw 
1-day average data. Then, we have calculated it for both weighted (see Appendix) 
and unweighted (Edelson et al.\ 2002) averages. In Fig.\ \ref{rms}, we show 
results for the unweighted averages. However, this method yields results similar 
to those using rebinned data and/or weighted averages, even if the values of the 
rms are somewhat different. For example, the hard-state data yield the weighted 
unrebinned fractional rms values of 0.68, 060, 0.53, 0.67, 0.52 and 0.42 in the 
6 consecutive channels of Fig.\ \ref{rms}.

During the hard state, the fractional 
rms of the ASM data decreases with energy. Then, there is a slight increase 
of the rms from the last ASM channel to the first BATSE one. This may be due to 
the presence of variable Compton reflection (e.g., Revnivtsev et al.\ 2001), 
which spectral component peaks at $\sim$30 keV (e.g., Magdziarz \& Zdziarski 
1995). Still, the trend of the rms decreasing with energy continues within 
the BATSE data, consistent with the pivot at $\ga 160$ keV (see Section 
\ref{highE}). We do not show the point for the 160--430 keV channel because of 
its large error. 

On the other hand, the rms increases with $E$ for either of the soft states. Interestingly, it is much higher for the second outburst, correlated with its substantially higher flux. The increase of the rms with energy in both cases can be interpreted as the high-energy tail being more variable than the blackbody component, as also found for Cyg X-1 (Z02). 

\subsection{Timing}
\label{timing}

\begin{figure}
\centerline{\psfig{file=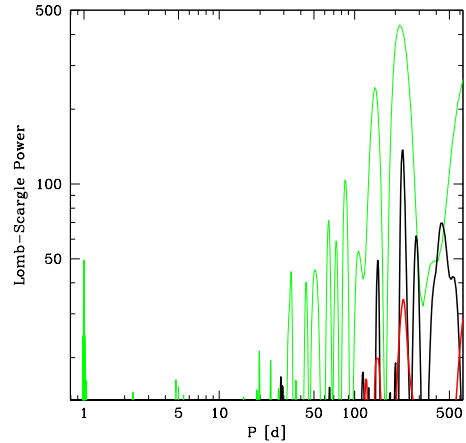,width=6.cm}} 
\caption{The Lomb-Scargle periodograms of the dwell 1.5--12-keV ASM data (green), the 40--70 keV BATSE data (black), and 1--20 keV \ginga/ASM data (red). The most significant peak in the respective data set is at 215, 227 and 230 d. The ASM peak at 1.0 d is instrumental, see Section \ref{timing}. The peak at $\sim$50 d probably corresponds to the precession period of the orbital plane of \xte. Only the power values above the 0.99 significance level for white noise in the ASM data (at 13.5) are plotted. The powers corresponding to this significance level for the other data sets are only slightly lower.
}
\label{period}
\end{figure}

We have studied periodograms (Lomb 1976; Scargle 1982; as implemented in Press 
et al.\ 1992) of the \xte/ASM (using the dwell data), BATSE and \ginga/ASM data. 
We find the most significant peaks in the periodograms at 215 and 227 d in the 
first two data sets limited to the hard state (MJD $<50800$), and at 230 d for 
the \ginga\/ data, see Fig.\ \ref{period}, confirming the results of Kong et 
al.\ (2002). (We note that their \xte/ASM peak power was much lower due to the 
use of the daily-averaged data; also, we refer the reader to that paper a 
thorough discussion of the sigificance levels.) We see that within their widths, 
all the peaks are compatible with being identical, $\simeq 220$ d. Kong et al.\ 
(2002) have attributed this period to the repetition time scale of the 
outbursts. However, visual inspection of the lightcurve, Fig.\ \ref{lc_all}, 
does not seem to confirm it. We see, e.g., that the first three BATSE outbursts 
were separated by $\sim$450 d, which apparently corresponds to the broad peak at 
433 d in the BATSE periodogram. To test it further, we have made the periodogram 
for the BATSE data constrained to those three outbursts, MJD $<49500$, and 
obtained results virtually identical to those in Fig.\ \ref{period}. Thus, we 
consider it likely that the 220 d period corresponds to the precession of the 
accretion disc in the system. This issue certainly deserves further study, 
which, however, is beyond the scope of this paper. 

On the other hand, no $\sim$220 d peak appears in either the entire \xte/ASM data set or the data constrained to any of the two recent soft outbursts. Thus, it appears that the first soft altered the accretion flow so significantly that the precession pattern changed. 

We have also searched for modulation corresponding to the orbital period, $\sim 
1.7$ d. We have not found any in the ASM data for either the hard (Fig.\ 
\ref{period}) or the soft states.  On the other hand, there is a clear peak at 
$\simeq 1.0$ d, which is also present in the ASM data for other sources, e.g., 
Cyg X-1 (Lachowicz et al., in preparation). It is an artefact related to the 
scheduling cycle of the \xte\/ observations, in  particular many pointings with 
a 1-day periodicity, e.g., during numerous AGN monitoring campaigns. Coupling of 
this periodicity with long-term secular changes introduces a powerful signal at 
$\sim$1 d.

\section{Comparison with Cyg X-1}
\label{cygx1}

As our results above show, there are many important similaraties and differences between \source\ and Cyg X-1. 

(i) Cyg X-1 shows no hysteresis. State transitions in that source follow the 
same trajectory in each direction on index-flux and index-index diagrams (Z02). 
The presence of hysteresis in \source\ is certainly related to its overall range 
of variability (and presumably the accretion rate) much wider than in Cyg X-1, 
both in terms of the fluxes and spectral indices. 

(ii) The 3--12 keV spectral index in the hard state correlates with the 
corresponding flux much less in \source\ (Fig.\ \ref{asm_index}c), than in Cyg 
X-1 (fig.\ 4b in Z02). However, the range of this $\Gamma$ in the hard state is 
almost the same in both source, $\sim$1.2--2.0, whereas the range of the flux is 
$\ga$10 times higher in \source. This then explains the weaker dependence. The 
correlation in Cyg X-1 was interpreted by Z02 as due to variable source geometry 
in the hard state, with a variable relative rate of cooling of the Comptonizing 
plasma. Such variable geometry is also likely to be present in \source, 
governing its spectral hardness, but with an additional hysteretic dispersion of 
the hard-state flux.

(iii) The 3--12 keV and 1.5--5 keV indices in the hard state are anticorrelated 
in \source\ (resembing some atoll sources), whereas they are strongly positively 
correlated in Cyg X-1 (fig.\ 5 in Z02). This differerence is likely to be 
related to the $T_{\rm bb}$ in the hard state being much higher in \source\ than 
in Cyg X-1. Indeed, Corongiu et al.\ (2003) obtain $T_{\rm bb}\sim 0.4$ keV in 
the hard state of \source, compared to $\sim 0.15$ keV in Cyg X-1 (Di Salvo et 
al.\ 2001; Frontera et al.\ 2001)\footnote{We note that Wilms et al.\ (1999) obtained $T_{\rm bb}\sim 0.15$ keV in \source\ based on \asca\/ data. However, those data cover a much narrower energy band than that of \sax, as well as they used used a model with a broken power law, whereas Corongiu et al.\ (2003) found only a single power law necessary. }. Then, the variability pattern of the 
blackbody emission (see Section \ref{hard}) affects the ASM band much more in 
\source\ than in Cyg X-1. 

(iv) The hard-state fluxes in bands from 20 to 160 keV are positively correlated 
with the local spectral indices in \source. The sign of the correlation changes 
above 160 keV, indicating a pivot around this energy. On the other hand, the 
hard-state pivot in Cyg X-1 is at a lower energy, $\sim$50 keV (Z02). As 
discussed by Zdziarski et al.\ (2003), a pivot at energies close to the 
high-energy cutoff (or higher) indicates the behaviour in which an increase of 
the power in soft photons providing seeds for thermal Comptonization is 
accompanied by an increase of the power supplied to the hot Comptonizing plasma. 
In the case of Cyg X-1, that power is close to constant in the hard state on 
long time scales (unlike \source). We also note that Cyg X-1 shows a pivot at 
$\sim$120 keV similar to that of \source\ when studied on hour time scales 
(Bazzano et al.\ 2003).

(v) Consistent with the pivot at high energies, the hard-state BATSE data show the rms decreasing with energy, unlike Cyg X-1, where it has a minimum and then increases again above 100 keV (Z02). 

(vi) In the soft state, the flux-index anticorrelation observed in Cyg X-1 was 
explained by Z02 as a result of constant luminosity in the soft component and 
varying luminosity of the hard, power-law spectrum. The stronger the power-law, 
the less visible the cutoff in the soft component and thus the harder spectrum.
In \source, this behaviour is also seen in the two soft outbursts (Fig\ \ref{asm_index}). However, the blackbody flux is also variable in \source\ (Fig.\ \ref{outbursts}).

(vii) The dependence of the rms on energy in the first soft outburst is similar in \source\ and in Cyg X-1 during its 1996 soft state (Z02). On the other hand, the second soft state shows the daily time scale rms higher than that of Cyg X-1 in any of its soft states. 

(viii) No modulation of X-rays with the orbital period is present in \source\ in any state, compared to the strong modulation of soft X-rays in the hard state of Cyg X-1 (e.g., Wen et al.\ 1999). This is again likely to be related to the different nature of the accretion flow in the two objects. No X-ray eclipses or dips in \source\ is due to inflow of matter being contained in the Roche-lobe overflow, unlike the case of the wind accretion in Cyg X-1. 

\section{Discussion and theoretical interpretation}
\label{theory}

\subsection{An accretion model for the \textbfit{RXTE}/ASM data}
\label{accretion}

Our results show that luminosity changes in \source\ are not simultaneous with 
changes of the spectral shape. Obviously, a physical model to describe this 
behaviour is desired. The data with best time coverage we have are from the 
\xte/ASM instrument, which, however, provides us only with information in three 
channels in the limited energy band of 1.5--12 keV. Given that, our theoretical 
model should be rather simple. Thus, we have developed a model of an accretion 
flow composed of an inner, hot, optically thin disc and an outer, cold, 
optically thick one with only two free parameters. The first of them is the 
accretion rate (assumed constant throughout the flow), which also represents the 
normalization of the spectrum, and the second, the transition radius between the 
hot and cold disc. 

\begin{figure*}
\centerline{\psfig{file=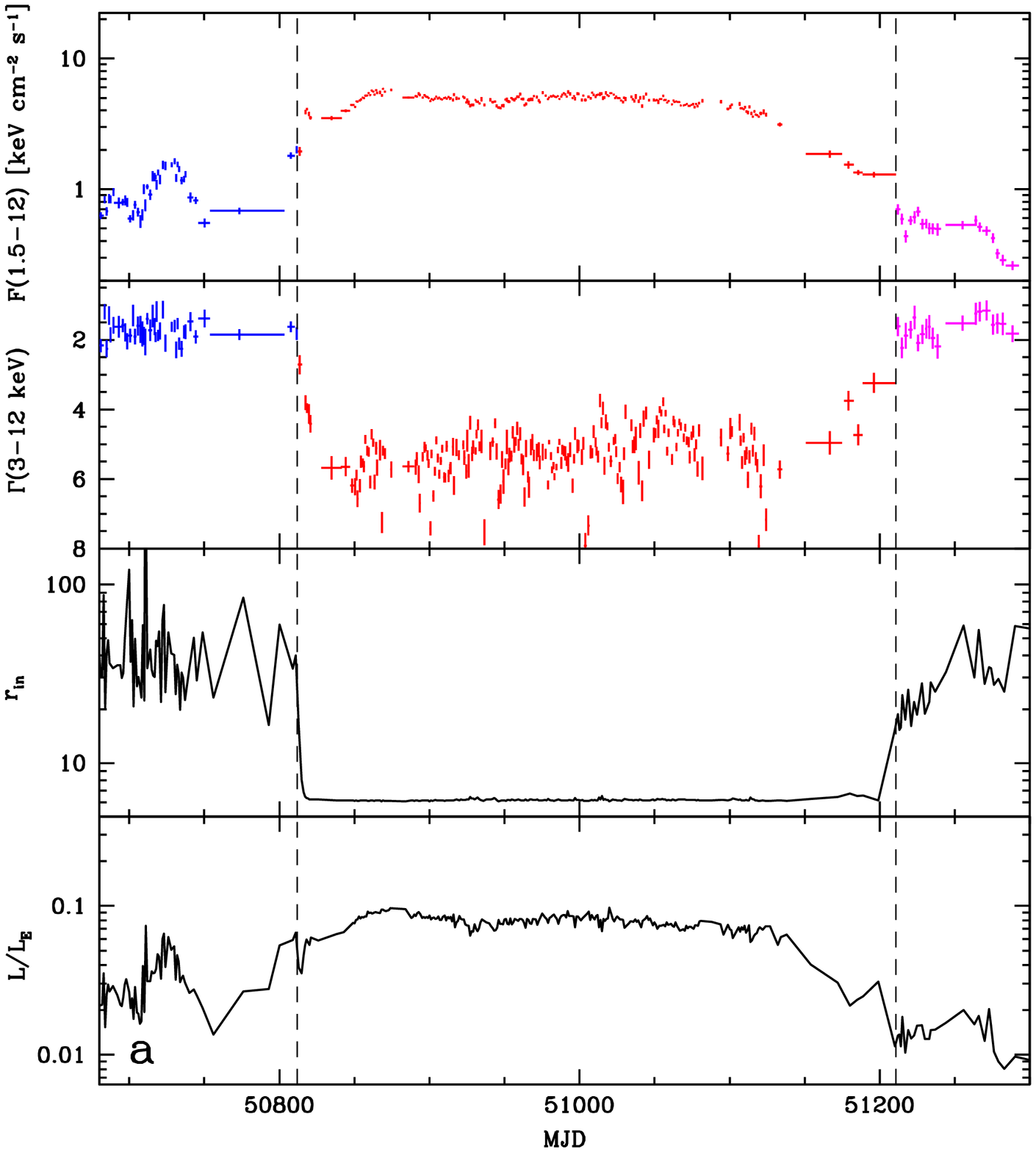,height=11.cm} \psfig{file=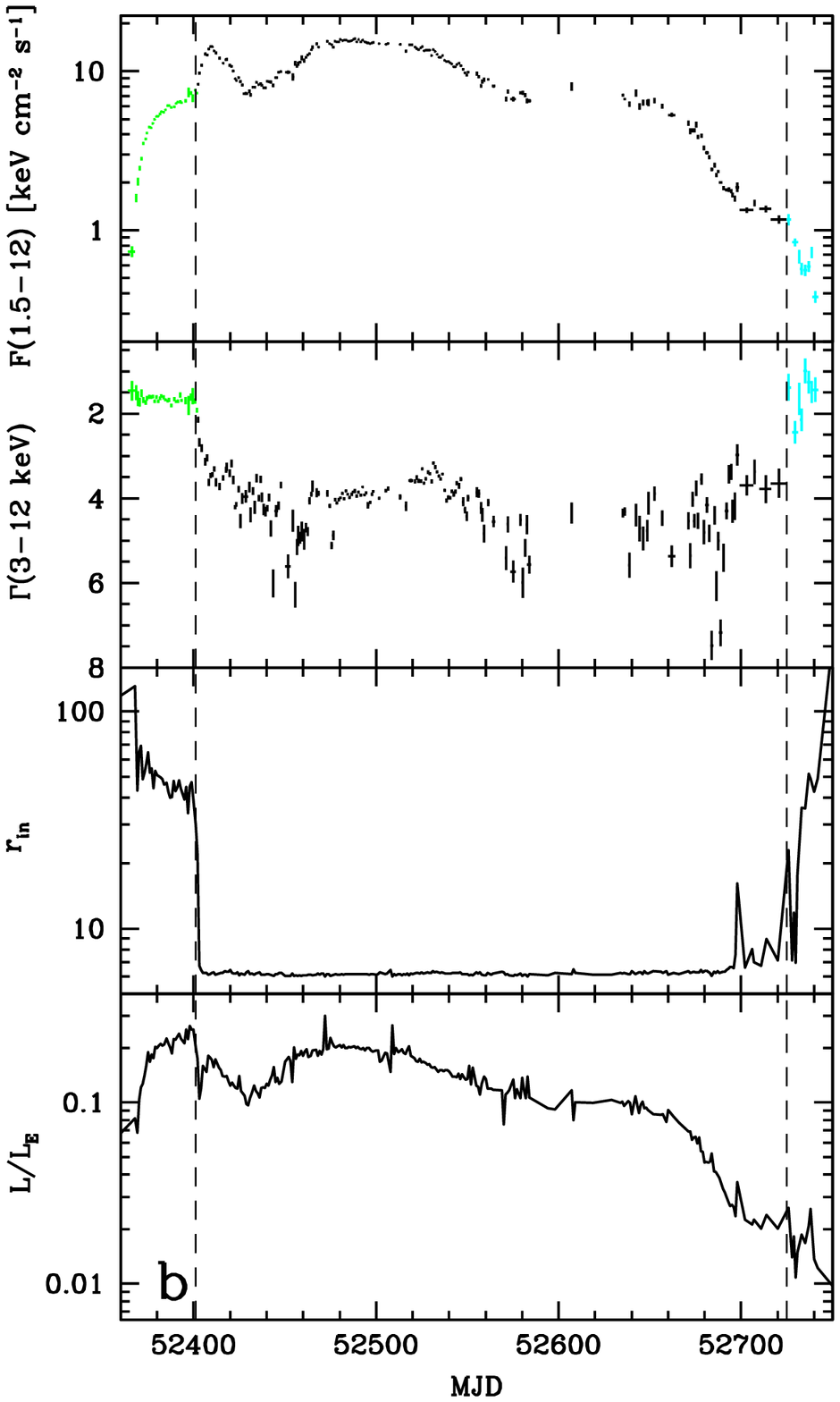,height=11.cm}}
\caption{The 1.5--12keV \xte/ASM flux and the 3--12-keV photon spectral 
index with the fitted values of $L/L_{\rm E}$ and $r_{\rm in}$ as a function of time during the two main outburst of \source. State transtions are marked by vertical lines. 
} 
\label{asm_model} 
\end{figure*}

The dimensionless accretion rate is defined in the standard way (e.g., Shakura \& Sunyaev 1973) as
\begin{equation}
\dot m \equiv {\dot M c^2\over L_{\rm E}},
\end{equation}
where $\dot M$ is the accretion rate and $L_{\rm E}$ is the Eddington luminosity,
\begin{equation}
\label{eddington}
L_{\rm E} \simeq 1.5\times 10^{38} \frac{M}{\msun} \;\; \mbox{erg s$^{-1}$}.
\end{equation}
Then $L=\eta \dot{m} L_{\rm E}$, where $\eta$ is an accretion efficiency.

We assume that an optically thick disc of Shakura \& Sunyaev (1973) extends down 
to a radius $r_{\rm in}$, expressed in the units of gravitational radius, 
$R_{\rm g}=GM/c^2$. The luminosity of the disc (in soft blackbody radiation) is
\begin{equation}
\label{ls}
L_{\rm S} = \frac{\dot{m} L_{\rm E} }{2  r_{\rm in}}.
\end{equation}
Below $r_{\rm in}$, we assume the flow becomes optically thin. This is intended to approximate optically thin accretion, e.g., cooling-dominated hot flows (Shapiro, Lightman \& Eardley 1976), ADAF (e.g., Narayan \& Yi 1996) or CDAF (e.g., Narayan, Igumenshchev \& Abramowicz 2000). We assume the luminosity of the optically thin flow is 
\begin{equation}
\label{lh}
L_{\rm H} = \frac{\dot{m} L_{\rm E} }{12 } - L_{\rm S},
\end{equation}
i.e., take the radiative efficiency of the hot disc equal to that of the cold one assuming no dissipation below $6r_{\rm g}$. This yields the accretion efficiency of $\eta=1/12$. 

The spectrum of the optically thick, outer, disc is approximated as a blackbody with the total luminosity of $L_{\rm S}$. The colour temperature, $T_{\rm bb}$, is calculated from the temperature of the Shakura-Sunyaev disc at $r_{\rm in}$, with the inner zero-torque boundary condition neglected for simplicity,
\begin{equation}
\label{tbb}
T_{\rm bb} = \kappa \left(
\frac{3 c^4 \dot{m} L_{\rm E} }{8 \upi \sigma G^2 M^2 r_{\rm in}^3}\right)^{1/4},
\end{equation}
where the colour correction of $\kappa=1.7$ is assumed (Shimura \& Takahara 1995). Since the inclination of \source\ is unknown, we consider here the angle-averaged emission.

For the hot disc spectrum we assume an exponentially cut off power law, which 
approximates the spectrum produced by unsaturated Comptonization of photons 
emitted by the disc. The power-law photon index is, 
\begin{equation}
\label{Gamma}
\Gamma = \frac{7}{3} \left(\frac{2 L_{\rm H}}{L_{\rm S}}\right)^{-\delta}
\end{equation}
(Beloborodov 1999) where $\delta$ depends on the temperature of seed photons, with $\delta \approx 1/6$ for stellar-mass black holes, and where we have assumed a half of the blackbody photons is Comptonized in the hot flow (and the other half escapes unscattered). The normalization of the power-law spectrum at the energy of the peak of the blackbody spectrum is calculated using eq.\ (7) in Zdziarski (1986). The energy of the cutoff in the power-law spectrum (depending on the electron temperature) is calculated by the normalization of that spectrum given by $L_{\rm H}$.

We note that our model allows for any $r_{\rm in}$ ($\geq 6$) for a given $\dot m$. Thus, it allows for either soft or hard state to take place at a given accretion rate, consistent with our observational result of hysteresis in \source.

We have implemented the above model into the {\sc xspec} package (Arnaud 1996). 
The \xte/ASM fluxes (in physical units and rebinned to $\ge 5\sigma$ 
significance, see Section \ref{data}) were converted to {\sc fits}-format files, 
assuming a diagonal response matrix. The above model was then fitted to the data 
assuming $d=8$ kpc, $M=10\msun$, and $N_{\rm H}=6\times 10^{21}$ cm$^{-2}$. The 
fit quality corresponded to the typical $\chi_\nu^2\la 1$ (although a higher 
$\chi_\nu^2$ was found for a number of days). 

\subsection{Model fits and bolometric luminosities}
\label{lum}

The results the fits with the above model are shown in Fig.\ \ref{asm_model}, in which we plot $L/L_{\rm E}$ ($=12\dot m$) instead of $\dot m$. The maximum model $L/L_{\rm E}$ (reached during the soft state) are then 0.10 and 0.30\footnote{Note that Maccarone (2003) estimated from the ASM data that the maximum bolometric flux during the second outburst was $1.4\times 10^{-7}$ erg cm$^{-2}$ s$^{-1}$, which is 2.4 times our ASM estimate. That overestimate is fully explained by the incorrect application of the ASM bolometric correction obtained for the hard state to the soft state (Maccarone, private communication).} (for the assumed $d$ and $M$) for the first and second outbursts, respectively. The corresponding global maximum of the daily ASM countrate was $73\pm 4$ s$^{-1}$ on MJD 52472, and the maximum ASM dwell countrate was $81\pm 9$ s$^{-1}$ on MJD 52472.9, i.e., only marginally higher than the daily one.  The maximum $L/L_{\rm E}$ in the hard state of 0.26 was achieved in the second outburst just before the state transition. The transition $L/L_{\rm E}$ corresponding to the transition to and out of the soft state were $\sim$0.07, $\sim$0.015, and $\sim$0.25, $\sim$0.01--0.025, for the first and second outburst, respectively. 

These ASM results can be compared to our estimated bolometric luminosities from 
the PCA/HEXTE. The highest (model) $L/L_{\rm E}$ achieved during the first 
outburst (at MJD 50865--50874, see Figs.\ \ref{lc_all}, \ref{outbursts}) is 0.12 
($F\simeq 2.3\times 10^{-8}$ erg cm$^{-2}$ s$^{-1}$), in good agreement with our 
model result (above) of 0.10. Then, the $L/L_{\rm E}$  some time before (MJD 
51201) and shortly after the soft-to-hard state transition (MJD 51222) is 0.034 
and 0.017 ($F\simeq 6.6$, $3.3\times 10^{-9}$ erg cm$^{-2}$ s$^{-1}$), 
respectively, in agreement with our ASM model results.

During the second outburst, the PCA/HEXTE data show the maximum in the hard 
state before the transition on MJD 52387--90 of $F\simeq 3.4\times 10^{-8}$ erg 
cm$^{-2}$ s$^{-1}$, $L/L_{\rm E}=0.17$, compared to the corresponding ASM model 
value of $\sim 0.2$. Then, the $L/L_{\rm E}$ during observations before [MJD 
52400, $\Gamma(3$--$6.4)\simeq 1.7$] and after the hard-to-soft transition [MJD 
52402, $\Gamma(3$--$6.4)\simeq 2.2$) are 0.14 and 0.21 ($F\simeq 2.8$, 
$4.1\times 10^{-8}$ erg cm$^{-2}$). On the other hand, the same values for 
observations just before [MJD 52727, $\Gamma(3$--$6.4)\simeq 2.7$] and after the 
soft-to-hard state transition [MJD 52734, $\Gamma(3$--$6.4)\simeq 2.0$) are 
0.031 and 0.020 ($F\simeq 6.1$, $3.9\times 10^{-9}$ erg cm$^{-2}$), with the 
transition itself occuring on MJD 52731 [$\Gamma(3$--$6.4)\simeq 2.2$] with 
$F=2.9\times 10^{-9}$ erg cm$^{-2}$, $L/L_{\rm E}=0.015$. The PCA/HEXTE global 
maximum was on MJD 52483 with $F\simeq 4.9\times 10^{-8}$ erg cm$^{-2}$ 
s$^{-1}$,  $L/L_{\rm E}=0.25$, in excellent agreement with our ASM model maximum 
of 0.30 on MJD 52472. Thus, the estimates of our ASM model, given all its 
simplifications, are in good agreement with the available broad-band data. The 
good agreement in the hard state also indicates the proper estimate of the 
high-energy cutoff energy in the ASM model.

Thus, the soft-to-hard state transitions in \source\ occur just around the 
average value of $\sim 0.02\ledd$ found for black-hole binaries by Maccarone 
(2003) provided $d\simeq 8$ kpc. Overall, $L$ follows changes in the observed 
1.5--12 keV flux,  but different values of $r_{\rm in}$ are possible for the 
same $L$. The variations of $r_{\rm in}$ follow those of $\Gamma(3$--$12\,{\rm 
keV})$, which due to the ratio $L_{\rm H}/L_{\rm S}$ being determined by $r_{\rm 
in}$ in our model. The inner disc radius assumes values of $\sim$10--100 and 
$\simeq$6 in the hard and soft state, respectively. 

\subsection{The outburst patterns}
\label{longterm}

The average accretion rate in \source\ is low enough for the accretion to be 
unstable, similarly to standard X-ray novae (e.g., Mineshige 1996), but with a difference of much shorter recurrence time scales. However, the long-term behaviour, see Fig.\ \ref{lc_all}, does not consist of regularly repeating outbursts, which suggests variability of the long-term average $\dot M$. 

The BATSE lightcurve appears consistent with an increase of that $\dot M$ from 
an epoch of strong outbursts every $\sim$500 d separated by periods of a very 
low flux (Asai et al.\ 1998) to the source becoming more persistent, with weaker 
peaks and brighter minima. To check it, we have calculated the average BATSE 
flux in the first of the two epochs, during MJD 48400--49500, which is $\sim 
1.6\times 10^{-9}$ erg cm$^{-2}$ s$^{-1}$. Based on the PCA data, we can 
estimate the bolometric correction as a factor $\sim 1.3$--1.5. A further 
correction is due to soft states appearing after the hard outbursts. Based on 
the \ginga/ASM data, the duration of soft states at that epoch was relatively 
short, and we estimate it as $\la 1.5$. Thus, our best estimate of the average 
flux is $\langle F\rangle \sim 3\times 10^{-9}$ erg cm$^{-2}$ s$^{-1}$, which 
corresponds to $\sim 0.015\ledd (d/8\,{\rm kpc})^2 (10\msun/M)$. Due to 
systematic errors (e.g., in the absolute calibration of the BATSE), the 
numerical coefficient above may be somewhat lower, $\sim$0.01. This $\langle 
L\rangle$ clearly corresponds to accretion in this source being strongly 
unstable. The average BATSE flux during the next epoch, MJD 49500--50800, 
characterized by weaker and more often hard otbursts with higher minimum fluxes, 
had then the average flux of about 1.6 times the preceding one, corresponding to 
$\sim 0.025\ledd (d/8\,{\rm kpc})^2 (10\msun/M)$.

A further increase lead to the first of the two recent soft states. We estimated the bolometric correction based on the PCA data, which yielded $\langle F\rangle\sim 2\times 10^{-8}$ erg cm$^{-2}$ s$^{-1}$ during its duration. This soft outburst apparently emptied the inner region of matter, and a $\sim$1000 d off state followed. The accumulated matter powered the last strong outburst, with $\langle L\rangle$ of its soft state of about twice the preceding one, and a $\sim$300 d duration. Averaging from the beginning of the first outburst (MJD 50800) until now MJD 53000 yields $\langle F\rangle\sim 10^{-8}$ erg cm$^{-2}$ s$^{-1}$, corresponding to $\sim 0.05\ledd (d/8\,{\rm kpc})^2 (10\msun/M)$, i.e., about three and two times that during the preceding epochs, MJD 48400--49500 and 49500--50800, respectively. This indicates variability of the accretion rate on the time scale of $\sim 10^3$ d.

Interestingly, the average luminosity, $\langle L\rangle$, during the initial 
outbursts observed by BATSE,  $\sim$(0.01--0.015)$\ledd (d/8\,{\rm kpc})^2 
(10\msun/M)$, is very similar to that of the hard state in a number of 
persistent sources, e.g., Cyg X-1, where $\langle L\rangle\sim 0.012\ledd 
(d/2\,{\rm kpc})^2 (10\msun/M)$ during 1991--94 (Z02), but also in 1E 
1740.7--2942 and GRS 1758--258 (Heindl \& Smith 1998). In the case of Cyg X-1, 
its stability as opposed to the instability of \source\ may be  related to the 
different sizes of accretion discs in those systems. The instabilities in 
\source\ certainly develop in the outer disc, whereas the accretion disc in Cyg 
X-1 may form out of the quasispherical wind relatively close to the black hole (Beloborodov \& Illarionov 2001; Illarionov \& Beloborodov 2001).

\subsection{State transitions}
\label{transitions}

The hysteresis in \source\ is obviously related to the existence of a range of $L$ at which two stable accretion solutions, optically thick and optically thin, exist. The state of the source in a given moment is then determined not solely by $L$ but also by the past history. The maximum possible luminosity of the hard state is larger than the minimum possible luminosity of the soft state. We could then expect the transitions to/out of the soft state to appear simply at the limiting luminosities.

The transition out of the soft state did take place in a narrow range of $L$
during the last two outbursts. Such a uniformity is also characteristic to other sources (Maccarone 2003). This luminosity is likely to represent the minimum possible soft state $L$, limited probably by evaporation (e.g., R\'o\.za\'nska \& Czerny 2000; Meyer, Liu \& Meyer-Hofmeister 2000). In this case, the past history determining the source behaviour is simply the presence of an optically-thick disc. 

However, it is less clear what determines the moment of collapse of the 
hard-state hot flow and its transformation into an optically-thick disc. The 
luminosity at which this happens varied by a factor of $\sim$3 between the two 
last outbursts of \source. The cause of this difference appears to be the 
previous history of the source. The transition to the first of the two 
soft states happened after several years of sustained hard-state activity. On 
the other hand, the second outburst followed a $\sim$1000 d of quiescence. One 
possibility is that the inner radius of the outer cold disc was much closer to 
the black hole when the $\dot M$ started to increase leading to the first soft 
outburst. Before the second one, the cold disc was initially much further out, 
and its buildup towards lower radii did not keep pace with the rapidly 
increasing local accretion rate. This, in turn, allowed a much higher luminosity 
in the hard state. Thus, the past history in the case of the hard-to-soft 
transitions appears to be not just existence a hot flow but also the state of 
the surrounding cold disc.

A way of testing the above hypothesis would be studying the evolution of the solid angle covered by the disc by measuring the strength of Compton reflection in the source. Indeed, an increasing strength of reflection with decreasing $r_{\rm in}$ is expected from consideration of energy balance (Zdziarski, Lubi\'nski \& Smith 1999), and consistent with the observed associated increase of the degree of relativistic broading of Fe K spectral features and the QPO frequency (Revnivstev et al.\ 2001). Nowak et al.\ (2002) noted also a correlation of the reflection strength with the X-ray flux before the first soft outburst. We would then expect different time histories of Compton reflection and the flux in the case of the hard state preceding the first and the second of the soft outbursts. 

\subsection{The nature of the hard state}
\label{hard}

\begin{figure}
\centerline{\psfig{file=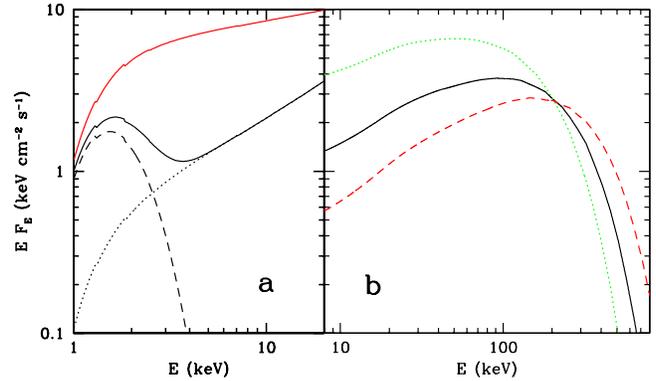,width=8.5cm}} 
\caption{(a) Model spectra explaining the anticorrelation between the 1.5--5 and 3--12 keV spectral indices seen in Fig.\ \ref{asm_index}(e). The upper and lower solid curves show total spectra corresponding to 100 and 10 per cent, respectively, of the soft blackbody flux cooling the hot plasma. The dotted and dashed curves correspond to the Comptonized emission and the unscattered blackbody emission, respectively, in the latter case. (b) Model spectra illustrating the variability pattern with the pivot at $\sim$200 keV and softening with increasing $L$. The solid curve shows an initial spectrum, which then undergoes perturbations leading to the spectra shown by the dashed and dotted curves. The dotted curve corresponds to an increase of the ratio of the power in soft seed photons incident on the hot plasma to the power in that plasma by a factor of 2, and, at the same time, an increase of the hot-plasma power by a factor of 2. In the the case of the dashed curve, the same quantities are decreased by factors of 2 and 3/2, respectively. See Section \ref{hard} for discussion.  
}
\label{patterns}
\end{figure}

We have found several interesting aspects of the variability of \source\ during 
the hard state. First of all, the strongest correlation in the \xte/ASM data is 
a negative one between the 1.5--5 keV and 3--12 keV average slopes, forming an 
upper horizontal branch in the colour-colour diagram, Fig.\ \ref{asm_index}(e). 
We find it can be very well explained in the framework of thermal 
Comptonization. As discussed, e.g., by Z02, the main factor determining the 
3--12 keV slope is the ratio between the power in soft photons incident on the 
Comptonizing plasma to the power supplied to that plasma, likely determined by a 
variable source geometry. A hardening of that slope corresponds to a reduction 
of that ratio. However, only a geometry-dependent fraction of the emitted soft 
flux is incident on the hot flow. The main variability pattern may consist of 
changing this fraction rather than the total soft flux. Then, a reduction of the 
incident fraction would both harden the Comptonized spectrum and decrease the 
point at which that spectrum intersects with the spectrum of the soft photons, 
as illustrated in Fig.\ \ref{patterns}(a). The change may be caused by a 
variable overlap of the cold disc by the hot flow (see fig.\ 15 in Z02).

In order to model this effect quantitatively, we have used a Comptonization 
model of Coppi (1999) (described in detail in Gierli\'nski et al.\ 1999), in 
which the plasma electron temperature is determined self-consistently from 
energy balance between the powers in the hot plasma and the incident seed 
photons. The  X-ray slope is determined by $T$ and the Thomson optical depth, 
$\tau$. We have assumed that there is a constant total power in the soft 
blackbody spectrum, but its fraction irradiating the hot plasma is variable, as 
discussed above. We assumed the soft photon spectrum is a disc blackbody with 
the maximum temperature of $kT_{\rm bb}=0.4$ keV (Corongiu et al.\ 2003), and 
$\tau=2$ (W02). 

At one extremum, all the blackbody photons irradiate the plasma (i.e., the cold 
disc is embedded in the hot flow), which power is assumed to be 5 times that in 
the blackbody photons. This spectrum, absorbed by $N_{\rm H}=6\times 10^{21}$ 
cm$^{-2}$, is shown in Fig.\ \ref{patterns} by the upper solid curve. Based 
on it, we calculate the fluxes in the ASM energy channels and  the effective 
spectral indices, obtaining $\Gamma(1.5$--$5\,{\rm keV})=1.48$ and 
$\Gamma(3$--$12\,{\rm keV})=1.67$. We now reduce the fraction of the blackbody 
photons incident on the hot plasma by a factor of 10 (e.g., due to a radial 
separation between the cold disc and hot flow). This strongly reduces the 
cooling rate, leading to both a spectral hardening of the power-law emission 
(dotted curve) and the appearance of the unscattered part of the blackbody 
(dashed curve). The resulting total spectrum (the lower solid curve in Fig.\ 
\ref{patterns}) has $\Gamma(1.5$--$5\,{\rm keV})=2.65$ and $\Gamma(3$--$12\,{\rm keV})=1.39$. The two points corresponding to those spectral solutions 
are joint by a heavy black line in Fig.\ \ref{asm_index}(e). We see this effect 
leads to a correlation with a slope identical to the best fit to the data. 

Second, we have found the high-energy spectrum softens with the increasing luminosity. In the thermal Comptonization model, which fits broad-band data for \source\ well (Z98, W02), the energy of the cutoff is directly related to $T$. As in W02, we consider a hot accretion model of Zdziarski (1998), parametrized by the Compton parameter, 
\begin{equation} y=4 \tau {kT \over m_{\rm e} c^2}, 
\end{equation} 
where $m_{\rm e} c^2$ is the electron rest energy. This parameter 
determines the slope, $\Gamma$, which is weakly dependent on the 
flux in the data, see Fig.\ \ref{asm_index}(c),  indicating a constant $y$. Then, $T\propto L^{-2/7}$ and $L^{-1/6}$ (where $L$ is the luminosity) in the advection and cooling dominated cases, respectively (Zdziarski 1998). This behaviour reflects more efficient cooling at higher accretion rates.

In order to test those predictions, we first fit a spectrum from a pointed 
observation (the \ginga/OSSE spectrum no.\ 1 of Z98) by a model including the 
intrinsic continuum due to thermal Comptonization. We now use a Comptonization 
model of Poutanen \& Svensson (1996), also used by W02, in which, in contrast to 
the model of Coppi (1999), $T$ is a free parameter. The resulting points are 
plotted by the rightmost asterisk in each of Figs.\ \ref{asm_index}(c) and 
\ref{batse_index}(a--d). We then reduce the luminosity by a factor of 5 and 
increase $T$ according to the above two dependences. We also consider the case 
with $L$ recuced by a factor of 10, and $T\propto L^{-2/7}$. The resulting three 
model points are shown by the three left-hand-side asterisks in each of Figs.\ 
\ref{asm_index}(c), \ref{batse_index}(a--d). We see in Fig.\ \ref{asm_index}(c) 
that the three points correspond indeed to almost the same X-ray slope as the 
original data. We also see that this model roughly reproduces the index 
increasing with the flux above 40 keV, Figs.\ \ref{batse_index}(b--c). It 
predicts an almost constant $\Gamma$ in the 20--70 keV range, not reproducing a 
slight increasing trend in the data, Fig.\ \ref{batse_index}(a). However, this 
increase of $\Gamma$ is solely due to the BATSE data preceding the \xte/ASM 
monitoring, for which we do not know the behaviour in the 3--12 keV range. Thus, 
this model provides a likely explanation of the behaviour seen in the hard-state 
BATSE data. 

Third, we have found a pivot point around $\sim$200 keV, Figs.\ 
\ref{batse_index}(c--d), which is similar to the variability pattern found in a 
number of Seyfert galaxies (Zdziarski et al.\ 2003). This implies that though a 
major driver of the variability is at low energies due to the variable flux of 
seed photons for thermal Comtonization, an increase/decrease of that flux is 
accompanied by a similar increase/decrease of the power in the hot 
Comptonizing plasma. Fig.\ \ref{patterns}(b) illustrates such a pattern, resulting in a pivot at 200 keV. We used the same models as those of fig.\ 14a of Z02, but adjusted their normalization as above. In that model, $T$ is self-consistently adjusted to varying irradiation by soft photons. 

Both above patterns, the decreasing $T(L)$, modelled for a constant X-ray 
slope, and pivoting, which changes the slope, act together. The pivoting may 
explain the softening of the 20--70 keV slope with the increasing flux, not 
explained by the $T(L)$ dependence. It also explains the disagreement of the 
trend predicted by the $T(L)$ model at energies above 160 keV, see the asterisks 
in Fig.\ \ref{batse_index}(d), where pivoting dominates. In fact, the variability pattern shown in Fig.\ \ref{patterns}(b) includes softening of the spectrum with increasing luminosity, as the increase is accompanied by a decreasing $T$ (see above and Z02). 

\subsection{Radio emission in the hard state}
\label{radio}

Altogether, thermal Componization explains well both the long-term spectral variability of \source\ in its hard state and its broad-band X\g-ray spectra (Z98, W02). On the other hand, synchrotron emission by nonthermal power law electrons in a jet was proposed to be responsible for its hard-state X\g-ray spectra (Markoff et al.\ 2003). This model is motivated by the presence of very good correlations between the radio and X-ray fluxes in a few bands above 3 keV found in \source\ (Corbel et al.\ 2003). However, the nonthermal synchrotron model for the X\g-ray spectra suffers from many problems (e.g., Zdziarski et al.\ 2003), in particular it requires the high-energy cutoff in the electron distribution to be both very sharp and highly fine-tuned. 

Our present results on spectral correlation add to the problems of this model. Especially, we have shown that the hard X-ray power law responds to the variability in the soft X-ray excess, with an increase of the relative strength of the excess associated with a hardening of the power law. In the nonthermal model, the power law extends down to the turnover frequency in the infrared, and it is not clear which mechanism would cause such behaviour (explained by thermal Comptonization cooling in our model). It is also unclear why the nonthermal power law should pivot around 200 keV. 

As discussed in Zdziarski et al.\ (2003), the likely origin of the radio-X-ray 
flux correlation is the formation of the radio jet out from energetic particles 
in the hot inner flow (i.e., the base of the jet being that flow), which also 
radiates X-rays. In particular, Heinz \& Sunyaev (2003) derived power law 
dependences between the jet radio flux and $\dot M$. On the other hand, the 
accretion luminosity also follows a power law dependence on $\dot M$, with the 
exponent $\geq$1 (e.g., Shapiro et al.\ 1976; Narayan \& Yi 1996; 
Zdziarski 1998). Consequently, there will be a power law dependence between the 
jet radio flux and the bolometric accretion luminosity. 

\begin{figure}
\centerline{\psfig{file=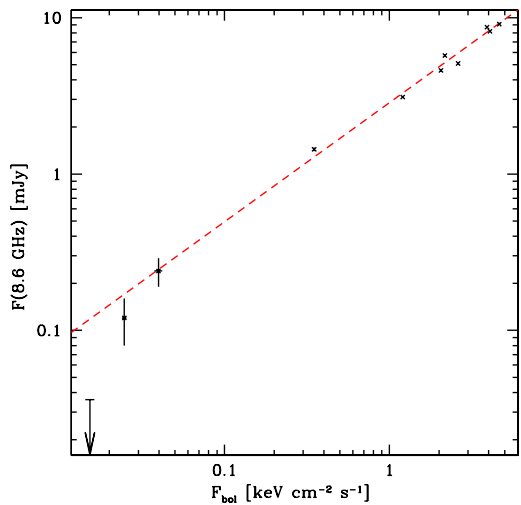,width=6.8cm}}
\caption{Correlation between the radio flux (Corbel et al.\ 2003) and the bolometric (estimated from our best-fit models of PCA/HEXTE data) X-ray flux in the hard state. The radio upper limit corresponds to MJD 51388. The dashed line shows the best fit to the data (excluding the upper limit). 
}
\label{xte_radio}
\end{figure}

In order to test this theoretical prediction, we plot in Fig.\ \ref{xte_radio} 
the dependence between the 8.6 GHz radio flux of Corbel et al.\ (2003) and the 
bolometric flux estimated by us from the PCA/HEXTE data. We have used only the radio measurements for which same-day X-ray measurements exist. With the current calibration of the PCA (see Section \ref{data}), we also obtain an X-ray detection on MJD 51388 (with the 3--16 keV flux of $5.0\pm 0.3\times 10^{-12}$ erg cm$^{-2}$ s$^{-1}$), for which Corbel et al.\ (2003) reported an upper limit. We obtain indeed an excellent correlation for all the points with the radio measurements. A least-square fit symmetric in both fluxes yields
\begin{equation}
F(8.6\,{\rm GHz})=2.79\pm 0.30 F_{\rm bol}^{0.79\pm 0.07},
\end{equation}
where we have included systematic errors with a fractional value (of 0.25) as implied by the condition of $\chi^2_\nu =1.0$, and the units are as in Fig.\ \ref{xte_radio}. Observationally, this systematic error is mostly due to systematic errors of calculating the bolometric $L$ (see Section \ref{data}), and the observations being not fully simultaneous. Interestingly, we see that the correlation breaks down at lowest fluxes as implied by the radio upper limit (which was not included in the fit). 

We note that our explanation of the correlation of Corbel et al.\ (2003) is different from that of Markoff et al.\ (2003), who attributed it to the X-ray flux and the optically-thin jet emission being the same power law of the 
nonthermal synchrotron. In our explanation, we do not require the jet optically-thin emission and the X-rays to lie on the same power law. 

\section{Conclusions}

We have found that the current optical/IR data imply that the distance to \source\ is $\ga 7$ kpc. Its most likely location is in the Galactic bulge. We also argue against the recently proposed distance of $\sim$15 kpc, in which case \source\ would be a halo object. 

Then, we have performed a comprehensive analysis of 16 years of the long-term 
variability of \source. We found the long-term $\langle L\rangle$ from this 
transient source has increased about threefold since 1991, corresponding to 
marked changes in the character of the outbursts. The epoch MJD 48400--49500 was 
characterized by a $\sim$500 d cycle of strong hard outbursts separated by deep 
quiescence and $\langle L\rangle\sim 0.015\ledd (d/8\,{\rm kpc})^2 (10\msun/M)$. 
The epoch MJD 49500--50800 was characterized by a persistent hard state and 
$\langle L\rangle$ about 1.5 times the above value. This increase lead then to a 
long soft state, a $\sim$1000 d queiescence, and the most recent $\sim$400 d 
very strong outburst. During this epoch, MJD 50800--53000, $\langle L\rangle\sim 
0.05\ledd (d/8\,{\rm kpc})^2 (10\msun/M)$. 

We have studied in detail hysteresis in this source. Especially interesting is the strong dependence of the flux of the hard-to-soft transition on the preceding behaviour of the hard/off state. The 1998 transition, which followed the long persistent hard-state period, took place at the flux about three times lower than the 2002 one, which followed the long queiescent epoch. We model the hysteretic variations by independent variability of the accretion rate and the inner radius of the cold disc. The maximum possible $\dot M$ in the hard state is higher than the minimum possible one in the soft state, but transitions from the hard to soft state can occur below that maximum. 

We have found a new X-ray spectral correlation in \source, between the relative strength of the soft excess (due to blackbody disc emision) in the hard state and the hardness of the main power law component. The effect is explained by cooling of the hot Comptonizing plasma by a variable fraction of the disc emission, probably related to a variable overlap of the disc by the hot flow. Interestingly, the correlation forms an upper horizontal branch in the colour-colour diagram, previously thought to be specific to neutron-star binaries. In contrast to Cyg X-1, we do not find any orbital modulation of the X-ray flux. 

We have also found a pivoting pattern of hard-state X-ray variability, with the pivot energy at $\sim$200 keV, and confirmed the existence of an anticorrelation between the bolometric flux and the high-energy cutoff, presumably related to the temperature of the Comptonizing electrons. The weak correlation of the X-ray slope and the flux (in contrast to the case in Cyg X-1) is explained by the wide range of the fluxes at which the the hard state appears, related to the transient nature of \source. 

Our results provide strong support to the thermal-Comptonization nature of X-rays in the hard state. We explain that the correlation between the X-ray and radio fluxes by the total power of the accretion emission being correlated with the power in outflowing particles forming the radio jet, which is, in turn, correlated with the radio flux.

\section*{ACKNOWLEDGMENTS}
This research has been supported by the KBN grants 5P03D00821, 2P03C00619p1,2, 
PBZ-KBN-054/P03/2001. We thank T. Belloni, C. Done, P. Grandi, P. Lachowicz, P. Lubi\'nski, T. Maccarone and P. \.Zycki for valuable and inspiring discussions, and the referee for valuable comments. We also thank S. Corbel for the radio data and 
acknowledge the use of data obtained through the HEASARC online service provided 
by NASA/GSFC, and, in particular, thank the \xte/ASM team for their data.

\appendix
\section{Weighted rms}

We consider a lightcurve of flux measurements, $F_i(t)$, $i=1,2,...,N$, each with a corresponding measurement standard error of $\sigma_i$. Its weighted average is,
\begin{equation}
\langle F\rangle = {\sum F_i/\sigma_i^2\over \sum 1/\sigma_i^2},
\end{equation}
which unbiased variance estimate is
\begin{equation}
\sigma^2_{\langle F\rangle} ={N\over N-1} {1\over \sum 1/\sigma_i^2}
\end{equation}
(e.g., Bevington \& Robinson 1992). The weighted average measurement-error variance is 
\begin{equation}
\langle \sigma_i^2 \rangle={N\over \sum 1/\sigma_i^2} = (N-1) \sigma^2_{\langle F\rangle}.
\end{equation}
The variance of this variance is (e.g., Brandt 1997),
\begin{equation}
\sigma^2_{\langle \sigma_i^2 \rangle}= {2\over N-1} \langle \sigma_i^2 \rangle^2.
\end{equation}

The unbiased estimate of the total flux variance (including both the intrinsic variability and measurement errors) is, 
\begin{equation}
\langle \sigma_{\rm t}^2\rangle= {N\over N-1} {\sum (F_i-\langle F\rangle)^2/\sigma_i^2\over \sum 1/\sigma_i^2},
\end{equation}
with its variance of
\begin{equation}
\sigma^2_{\langle \sigma_{\rm t}^2 \rangle} ={2\over N-1} \langle \sigma_{\rm t}^2 \rangle^2.
\end{equation}
The intrinsic source variance is,
\begin{equation}
\langle \sigma^2\rangle =\langle \sigma_{\rm t}^2\rangle-\langle \sigma_i^2\rangle,
\end{equation}
where its variance is the sum of the variances of the two factors,
\begin{equation}
\sigma^2_{\langle \sigma^2\rangle}= \sigma^2_{\langle \sigma_{\rm t}^2 \rangle} + \sigma^2_{\langle \sigma_i^2 \rangle}.
\end{equation}

The fractional variability is then,
\begin{equation}
R={\langle \sigma^2\rangle^{1/2} \over \langle F\rangle}.
\end{equation}
Considering contributions to its standard deviation from the three variances of its components, the relative error on $R$ is
\begin{equation}
{\sigma_R\over R}= {1\over [2(N-1)]^{1/2}} \left[ {\langle \sigma^2_{\rm t}\rangle^2+ \langle \sigma_i^2\rangle^2 \over (\langle \sigma^2_{\rm t}\rangle- \langle \sigma_i^2\rangle)^2} + {\langle \sigma_i^2\rangle^2 \over \langle F\rangle^2} \right] ^{1/2}.
\end{equation}
Note that when the average measurement error is small, the relative error is simply $\simeq (2N)^{-1/2}$. See Edelson et al.\ (2002) for similar calculations for unweighted averages, where, however, they neglected contributions to the error on $R$ from the variance of the average flux and average measurement error. 

\label{lastpage}

\end{document}